\documentclass[aps,pra,twocolumn,superscriptaddress,longbibliography, nofootinbib, 10pt]{revtex4-2}
\usepackage[T1]{fontenc}
\usepackage[utf8]{inputenc}
\usepackage{color}
\usepackage{svg} 
\usepackage{xcolor}
\usepackage{tikz}
\usepackage{graphicx}
\usepackage{babel}
\usepackage{amsmath,amssymb}
\usepackage{dsfont}
\usepackage{physics}
\usepackage{comment}
\usepackage[normalem]{ulem}
\usepackage[unicode=true,pdfusetitle,
bookmarks=true,bookmarksnumbered=false,bookmarksopen=false,
 breaklinks=true,pdfborder={0 0 0},pdfborderstyle={},backref=false,colorlinks=true,allcolors=blue] 
 {hyperref}
\usepackage[nolist]{acronym}
\usepackage{multirow}

\renewcommand{\epsilon}{\varepsilon}

\newcommand{\IP}[1]{\text{IP}\left(#1\right)} 
\newcommand{\IS}[1]{\text{IS}\left(#1\right)} 
\newcommand{\X}[0]{x} 
\renewcommand{\d}[0]{\textrm{d}} 

\DeclareMathOperator{\Poly}{Poly}

\newcommand{\eviden}{Eviden Quantum Lab, 78340 Les Clayes-sous-Bois, France}

\begin{document}

\title{Approximate combinatorial optimization with Rydberg atoms: the barrier of interpretability}

\author{Christian de Correc}
\email[]{c.de-correc@eviden.com}
\author{Thomas Ayral}
\author{Corentin Bertrand}
\affiliation{\eviden}

\begin{abstract}
Analog quantum computing with Rydberg atoms is seen as an avenue to solve hard graph optimization problems, because they naturally encode the \ac{MIS} problem on \ac{UD} graphs, a problem that admits rather efficient approximation schemes on classical computers.
Going beyond UD-\ac{MIS} to address generic graphs requires embedding schemes, typically with chains of ancilla atoms, and an interpretation algorithm to map results back to the original problem.
However, interpreting approximate solutions obtained with realistic quantum computers proves to be a difficult problem.
As a case study, we evaluate the ability of two interpretation strategies to correct errors in the recently introduced \ac{CL} embedding.
We find that one strategy, based on finding the closest embedding solution, leads to very high qualities, albeit at an exponential cost. 
The second strategy, based on ignoring defective regions of the embedding graph, is polynomial in the graph size, but it leads to a degradation of the solution quality which is prohibitive under realistic assumptions on the defect generation.
Moreover, more favorable defect scalings lead to a contradiction with well-known approximability conjectures. 
Therefore, it is unlikely that a scalable and generic improvement in solution quality can be achieved with Rydberg platforms---thus moving the focus to heuristic algorithms.
\end{abstract}

\maketitle

\begin{acronym}
    \acro{MIS}{Maximum Independent Set}
    \acro{MWIS}{Maximum Weighted Independent Set}
    \acro{UDMIS}{Unit-Disk Maximum Independent Set}
    \acro{UD}{Unit Disk}
    \acro{IS}{Independent Set}
    \acro{DoS}{Density of States}
    \acro{QAC}{Quantum Adiabatic Computation}
    \acro{CL}{Crossing Lattice}
    \acro{PTAS}{Polynomial-Time Approximation Scheme}
    \acro{fPTAS}{fully-\acl{PTAS}}
    \acro{QUBO}{Quadratic Unconstrained Binary Optimization}
\end{acronym}

\section{Introduction} \label{sec:intro}
\acresetall

\begin{figure}
    \centering
    \includegraphics[width=1\linewidth]{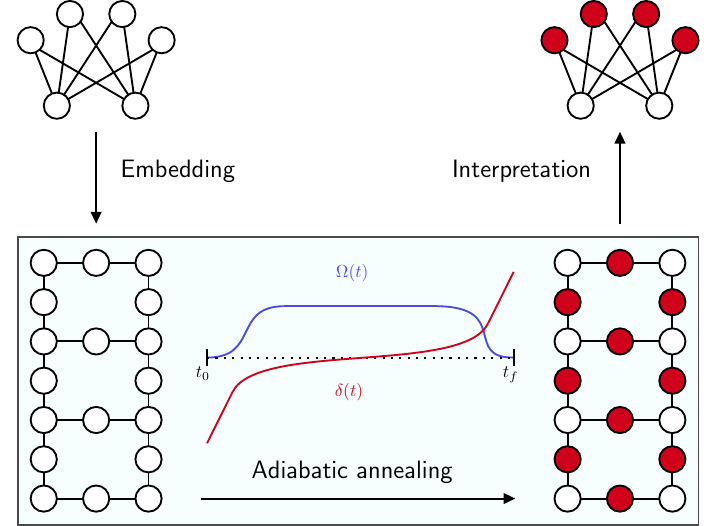}
    \caption{General procedure to solve \ac{MIS} on a generic graph by the means of an adiabatic evolution on a \ac{UD} graph. Red: selected vertices after the \ac{UD}-\ac{MWIS} exact resolution.
    }
    \label{fig:worflow-annealing-for-MIS}
\end{figure}

The \ac{UD}-\ac{MWIS} problem is the native use case of Rydberg quantum computing platforms due to its intimate link with the Rydberg blockade mechanism~\cite{pichler-2018-computational, pichler-2018-experimental, ebadi-2022, kim-2022-3d-rydberg-wires, Oliveira-2024, cazals-2025-identifying-hard-native-instances}. 
However, even if it is a NP-complete problem~\cite{karp-1972-reductibility-combinatorial-problems, garey-1978-MIS-strongly-NP-hard}, finding approximate solutions to \ac{UD}-\ac{MWIS} is relatively easy with classical algorithms~\cite{Erlebach-PTAS-UD-MWIS, matsui-1998-approximating-MIS-on-UDG, Das-2015-approximating-MIS-on-UDG, Nandy-2017-approximating-MIS-on-UDG, bonamy-2018-eptas-UBMIS}: one can get arbitrarily close to the optimal solution in a time polynomial in the graph size.
It raises the bar for reaching a putative quantum advantage over classical methods~\cite{serret-2020-benchmark}.
This has spurred proposals to tackle the general \ac{MIS} problem---which is much harder to approximate~\cite{Hastad-1999-hardness-approximating-MIS, zuckerman-2006-hardness-approximating-MIS}---with Rydberg platforms. 
This is done by mapping (or "embedding")  the general MIS problem to a larger \ac{UD}-\ac{MWIS} problem~\cite{pichler-2018-computational, dalyac-2023-phd-thesis, nguyen-2023}.
In principle, an optimal solution of this \ac{UD}-\ac{MWIS} problem is then obtained with a quantum computer, for instance via an adiabatic evolution~\cite{farhi-2001-quantum, albash-2018-adiabatic-computing}, and can be interpreted to produce a solution of the original MIS problem as schematized in Fig.~\ref{fig:worflow-annealing-for-MIS}.

In this work, we investigate the  robustness of embedding methods to imperfections in the \ac{UD}-\ac{MWIS} solution: imperfect \ac{UD}-\ac{MWIS} solutions, which are bound to appear in quantum algorithms due e.g to too short annealing times~\cite{Schiffer-2024-Circumventing-exp-runtimes, miessen-2024-defect-density-annealing, bombieri-2024-CL-gap}, cannot be directly mapped back to MIS solutions.
We study two complementary interpretation methods, akin to error mitigation methods, that handle these imperfections to reconstruct approximate MIS solutions.

More specifically, we focus on the recently introduced embedding called the \ac{CL}~\cite{nguyen-2023}, which comes with a quadratic overhead in the number of vertices.
While various interpretation algorithms are possible, we focus on two opposing strategies.
The first one, introduced as the \textit{distance} strategy in Sec.~\ref{sec:method}, arguably yields \ac{UD}-\ac{MWIS} solutions closer to the optimal one, but is computationally expensive.
In Sec.~\ref{sec:interpretability}, we give evidence that this strategy is limited by features of the \ac{CL} solution space.
The second strategy, called \textit{deselection} strategy, proves to be inexpensive but yields lower quality results (Sec.~\ref{sec:deselection}).
In Sec.~\ref{sec:discussion} we use our numerical results to derive scaling laws and discuss their implications in terms of the approximability of the MIS solution with Rydberg platforms, before concluding in Sec.~\ref{sec:conclusion}.

\section{Method and definitions}\label{sec:method}

Let $G = (E, V, W)$ be a simple finite weighted graph with set of vertices $V$, set of edges $E$ and weight function $W:V \longrightarrow \mathbb{R}^*_+$. 
A subset $S \subset V$ is said to be an \ac{IS}, or $S \in \IS{G}$, when no pair of vertices in $S$ is adjacent in $G$. 
The \ac{MWIS} problem amounts to selecting vertices that form an \ac{IS} with maximal total weight.
It is equivalent to finding a ground state of the Ising Hamiltonian
\begin{equation}
\label{eq:hamiltonian}
    \mathcal{H}_{\text{MWIS}} 
    = 
    - \delta \sum_{i \in V} W(i) \; n_i
    + U \sum_{\substack{(i,j) \in E }} n_i n_j
\end{equation}
where $n_i = 1$ if the vertex $i$ is selected and 0 otherwise. 
Here, $\delta > 0$ is an energy unit and $U \gg \delta$ is the interaction energy which implements the independence constraint in the limit $U \rightarrow +\infty$.
On \ac{UD} graphs, this Hamiltonian can be implemented with Rydberg atoms, thanks to the Rydberg blockade that acts as a repulsive interaction between neighboring atoms in the excited Rydberg state~\cite{pichler-2018-computational, urban-2009-rydberg-blockade}.
For the rest of this paper we assume a perfect implementation of the independence constraint, therefore ignoring the impact of the Rydberg interaction long range tail.
We also set the energy scale to $\delta = 1$ and omit the notation of $\delta$ from now on.

On arbitrary graphs, the \ac{MWIS} problem can be reduced to \ac{UD}-\ac{MWIS} by using an \textit{embedding}.
In this work, for simplicity, we focus on \ac{MWIS} embeddings of the \ac{MIS} problem, that is when $W = 1$ for the embedded graph.
We don't expect the generalization to \ac{MWIS} to significantly change our conclusions.
Formally, we define an embedding as a polynomial-time algorithm, which maps an instance $G$ of \ac{MIS} onto an instance $G_\mathrm{em}$ of \ac{UD}-\ac{MWIS} called \textit{embedding graph}, and an injective function $f:\IS{G}\longrightarrow\IS{G_\mathrm{em}}$ called \textit{embedding map}.
The map $f$ should preserve the weight ordering, i.e. for every pair $S, T \in \IS{G}$,
\begin{equation} \label{eq:weight-ordering-preservation}
    W(S) \geq W(T) \implies W_\mathrm{em}(f(S)) \geq W_\mathrm{em}(f(T)),
\end{equation}
where $W$, $W_\mathrm{em}$ are the weight profiles on $G$, $G_\mathrm{em}$.
Being injective, $f$ admits an inverse $f^{-1}$ defined on $f(\IS{G})$ and called the \textit{interpretation map}.
Finally, we require $f^{-1}$ to map any \ac{MWIS} of $G_\mathrm{em}$ to an \ac{MWIS} of $G$.

Being given an embedding scheme with some $G$, $G_\mathrm{em}$ and $f$, any configuration $S_\mathrm{em} \in f(\IS{G})$ is said to be \textit{interpretable}, since $f^{-1}(S_\mathrm{em})$ is well-defined and is an \ac{IS} for $G$.
Conversely, any configuration $S_\mathrm{em} \notin f(\IS{G})$ is said to be \textit{non-interpretable}.
Non-interpretable states arise due to the approximate resolution of \ac{UD}-\ac{MWIS} on $G_\mathrm{em}$, and they require some post-processing to be interpreted as configurations of the initial problem on $G$.

One possible strategy, which we name \emph{distance} strategy, consists in interpreting each configuration of the \ac{CL} as the nearest interpretable configuration, with regard to the Hamming distance.
This distance, denoted $d$, can be seen as the minimum number of bits to flip to reach an interpretable state.
We take $d$ both as a measure of the difficulty to apply this strategy and as a quantity of interest to study the \ac{CL} solution space.
A large $d$ indicates that a larger part of the CL must be explored to find the nearest interpretable configuration, which correlates to a larger computational difficulty.
As shown in Appendix~\ref{app:algo_NIS}, applying this distance strategy is equivalent to solving a \ac{QUBO} problem closely related to an \ac{MWIS} problem on a subgraph of the original graph, showing that it can be a difficult task.
Note that our definition of $d$ and most subsequent discussions readily apply to other embeddings schemes than the \ac{CL}, such as the ones discussed in Sec.~C of the Supplemental Material~\cite{supp}.

\begin{figure}
    \centering
    \includegraphics[width=1.0\linewidth]{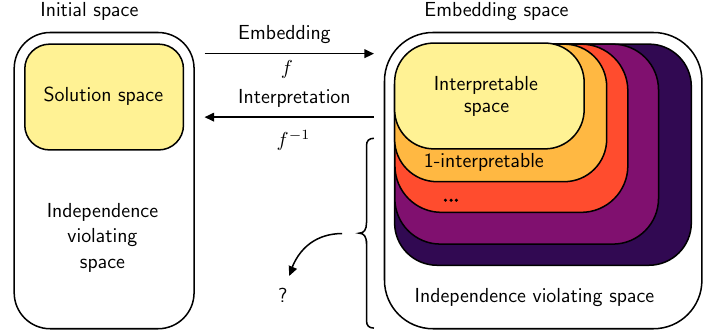}
    \caption{ 
    Structure of the initial and embedding solution spaces for a given embedding scheme, with sub-spaces of solutions characterized by their Hamming distance $d = 0, 1, \dots$ to the nearest interpretable solution.
    }
    \label{fig:layered-solution-space}
\end{figure}

Building upon these definitions, any state with a distance (at most) $d$ to the nearest interpretable state is thus called (at most) $d$-\textit{interpretable}.
With this definition, interpretable states are characterized by $d = 0$, and higher values of $d$ are associated to states that are harder to interpret.
As illustrated in Fig.~\ref{fig:layered-solution-space}, this gives a layered structure to the embedding space $\IS{G_\mathrm{em}}$, with easy-to-interpret solutions at the center, and increasingly difficult configurations around.

Throughout this article, the quality of approximate solutions is evaluated by the \textit{approximation ratio}, as is standard in optimization problems.
For any $S \in \IS{G}$, its approximation ratio $r \in [0, 1]$ is defined as the factor by which its weight deviates from the optimal value:
\begin{equation}
\label{eq:def-approximation-ratio}
    r
    = { 
        \displaystyle 
        \sum_{i \in S}W(i) 
        \over 
        \displaystyle 
        \max_{S' \in \IS{G}} 
        \sum_{i \in S'}W(i)
        }.
\end{equation}

\section{Distance strategy and interpretability}\label{sec:interpretability}

We perform a numerical study of $d$-interpretable states at low energies, obtained thanks to an exact \ac{MWIS} solver~\cite{Liu-2023-GNT}.
In Sec.~\ref{sec:path-embedding}, we begin by applying our methodology on a simple class of graphs called \textit{paths graphs} that can be seen as elementary components of \ac{CL} graphs.
In Sec.~\ref{sec:full-embeddings} we extend our study of the distribution of $d$ to \ac{CL} graphs.

\subsection{Path embeddings} \label{sec:path-embedding}

A core ingredient in \ac{CL} embedding is the class of \textit{path graphs}, which are \ac{UD} graphs used to embed the state of a single-vertex graph.
As a preliminary, we study them separately before looking at the \ac{CL}.

\begin{figure}
    \centering
    \includegraphics[width=1.0\linewidth]{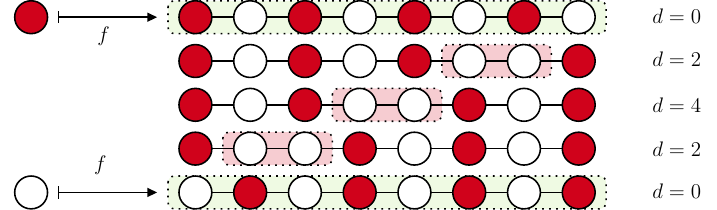}
    \caption{ 
    Path embedding of a single-vertex graph into two interpretable configurations \textit{(green highlight)}, with selected vertices in red. 
    Due to domain walls \textit{(red highlight)}, some configurations with a low energy have a high Hamming distance $d$  to the closest interpretable configuration.
    }
    \label{fig:MISs-even-path}
\end{figure}

For any $N \geq 0$, the (unweighted) path graph $P_N$ is defined as $N$ vertices $v_1, \dots, v_{N}$ and edges $(v_i, v_{i+1})$ for all $i = 1, \dots, N-1$.
Of particular interest is the case of the even-length path $P_{2N}$.
As shown in Fig.~\ref{fig:MISs-even-path}, in the unweighted case, the \ac{MIS} of $P_{2N}$ has $N$ selected vertices (in red) and a $(N+1)$-fold degeneracy i.e. number of configurations at the same energy.
In this context, we call \textit{domain wall} the presence of two adjacent unselected vertices.
In the unweighted $P_{2N}$, only two \ac{MIS}s have no domain wall; they are said to be \textit{antiferromagnetic}.

Antiferromagnetic configurations can be used to embed in $P_{2N}$ a single-vertex graph $G = (V=\{v\}, E=\varnothing)$.
The embedding map $f$ is defined such that $f(\{v\})$ is the antiferromagnetic state of $P_{2N}$ where $v_1$ is selected, and $f(\varnothing)$ is the other one.
The fulfillment of Eq~\eqref{eq:weight-ordering-preservation}, and thus the optimality of the embedded \ac{MIS} of $G$, requires to energetically penalize  domain walls, thus lifting the degeneracy between the two antiferromagnetic states.
To do so, we follow Ref.~\cite{nguyen-2023}, applying a weight profile $W_w$, parametrized by a \textit{weight bias} $w \in (0, 1/2)$, which reads
\begin{equation} \label{eq:weight-profile-chain}
    \begin{cases}
    W_w(v_1) & = 1/2 + w, \\
    W_w(v_{2N}) & = 1/2 - w,\\
    W_w(v_i) & = 1 \qquad \text{for} \quad 1 < i < 2N.\\
    \end{cases}
\end{equation}
Note that while we also discuss in Appendix~\ref{app:approx-ratio} the theoretical case $w \ge 1/2$, this paper uses $w < 1/2$ everywhere else.

\begin{figure}
    \centering
    \includegraphics[width=1.0\linewidth]{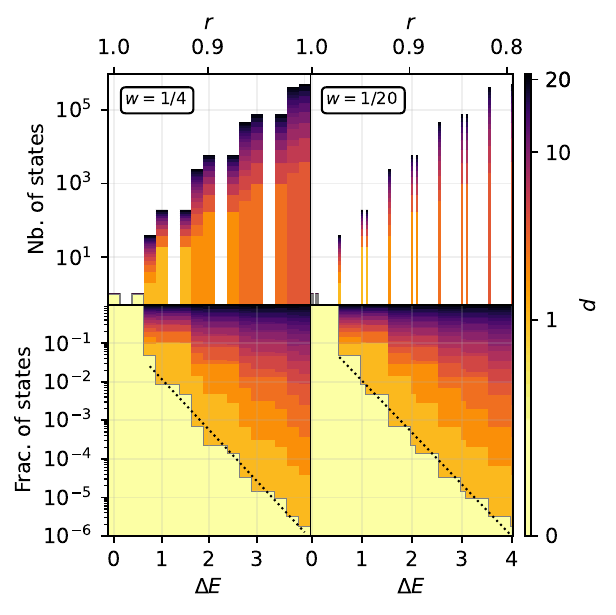}
    \caption{\textit{(Top)} Low-energy \ac{DoS} in the path graph $P_{40}$ with $1/w \in \{4, 20\}$.
    On the left, the two interpretable states have a gray outline to enhance readability.
    \textit{(Bottom)} 
    Distribution of $\tau_d$ in increasing energy windows, with an exponential decrease of $\tau_0(E)$ (dotted line).
    }
    \label{fig:DoS-one-weighted-chain}
\end{figure}

We now look at the corresponding \ac{DoS}. 
The number of states at each energy is displayed in the top panel of Fig.~\ref{fig:DoS-one-weighted-chain} as a function of the  energy
\begin{equation}
    \Delta E = E - E_{\text{MWIS}},
\end{equation}
and with a color scheme to indicate the distribution of the different values of $d$ at each energy.
Note that the energy is linearly related to the approximation ratio $r \in [0, 1]$ defined in Eq.~\eqref{eq:def-approximation-ratio}, with $r \simeq 1$ for the best approximate solutions.
A closed formula for this \ac{DoS} is derived in App.~\ref{app:DoS-paths-graphs} using graph combinatorics~\cite{Levit-Mandrescu-2003, arocha-1984-propriedades}, while the value of $d$ is computed algorithmically.
We observe that, even at energies slightly above the threshold $\Delta E = 1/2+w$, there is a high number of non-interpretable states with high values of $d$..
This can be interpreted in terms of domain walls, introduced earlier.
As shown in Fig.~\ref{fig:MISs-even-path}, the presence of only one domain wall in $P_{2N}$ can lead to a high Hamming distance, up to $d = N$.
There is thus an energy threshold around the domain-wall energy penalty $\Delta E = 1/2 + w$, below which there are only interpretable states, and above which non-interpretable states coexists with interpretable states.

To interpret this data in terms of probability of success after an adiabatic evolution, we look at the fraction $\tau_d(E)$ of at most $d$-interpretable states within the energy window $[E_{\text{MWIS}}, E]$.
Formally,
\begin{equation}
\label{eq:def-fraction-tau-path-graph}
    \tau_d(E) = \frac{\int_{E_{\rm MWIS}}^E \dd \omega \sum_{d' \le d} \rho_{d'}(\omega)}{\int_{E_{\rm MWIS}}^E \dd \omega \rho(\omega)}
\end{equation}
where $\rho$ is the total \ac{DoS} and $\rho_{d'}$ is its restriction to $d'$-interpretable states, with $\rho = \sum_{d'} \rho_{d'}$.
This measures the probability to get an easy-to-interpret state from random picking below energy $E$, and is shown in Fig.~\ref{fig:DoS-one-weighted-chain}~\textit{(bottom)}.
In the regime $\Delta E < 1/2 + w$, all states are interpretable since no domain walls are allowed.
Above this threshold, non-interpretable states start to appear, and for $1/2 + w < \Delta E \lesssim N/4$, $\tau_d(E)$ decreases exponentially for all values $d \ll N$.
This observation can be accounted for by the high degeneracy of domain walls, which is expected to grow exponentially with the energy.
Note that this behavior is consistent when $w$ varies, including when $w \ll 1$ (Fig.~\ref{fig:DoS-one-weighted-chain}, right column), for which the energy of a configuration is $\simeq l/2$ with $l$ the number of defects in the chain.

We conclude that in the path graph embedding, states of energy above the threshold $\Delta E = 1/2 + w$ become difficult to interpret with the distance strategy, and exponentially more so as their energy increases.
Quantum adiabatic computation should therefore aim at energies no much greater than this threshold.
In the next section, we extend this result to \ac{CL} graphs.

\subsection{Crossing lattice embedding} \label{sec:full-embeddings}

\begin{figure}
    \includegraphics[width=1.0\linewidth]{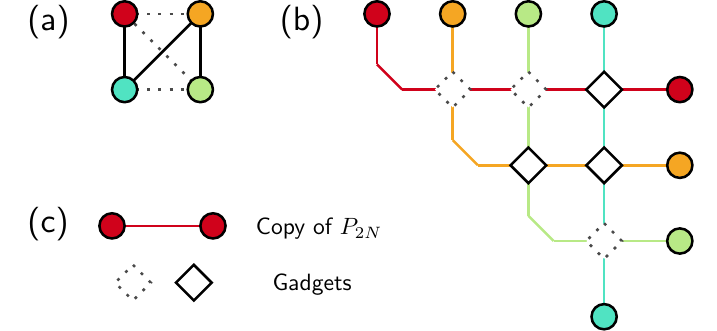}
    \caption{
    \textit{(a)} Initial graph with $N = 4$ vertices, $\vert E\vert = 3$ edges \textit{(plain lines)} and $N(N-1)/2 - \vert E \vert = 3$ pairs of non-adjacent vertices \textit{(dashed lines)}.
    \textit{(b)} Corresponding \ac{CL} embedding, with $N$ even-length paths used to embed the $N$ initial vertices and interconnected by $N(N-1)/2$ gadgets. Colors identify the mapping between initial vertices and paths in the \ac{CL}. \textit{(c)} Legend, with two types of gadgets subgraphs representing initial edges \textit{(plain diamond)} or their absence (\textit{dotted diamond)}.
    }
    \label{fig:schema-CL}
\end{figure}

Building upon the idea of the path graph embedding, the \ac{CL} scheme~\cite{nguyen-2023} starts by embedding the $N$ vertices of an initial graph $G$ into $N$ copies of $P_{2N}$ (or $P_{2N+2}$).
These paths are then intertwined along a square lattice grid, and interconnected by two types of subgraphs called \textit{gadgets} to enforce constraints caused by the presence or absence of edges between pairs of vertices in $G$.
A schematic example is displayed in Fig.~\ref{fig:schema-CL}.

In what follows, we examine the \ac{DoS} of \ac{CL} embeddings of unweighted random Erdős–Rényi–Gilbert graphs~\cite{gilbert-1959-random-graphs, correc-2025} with edge probability $p = 1/2$.
We generate $20$ unweighted graphs for each size $N = 3, \dots, 7$, and study their \ac{CL} embedding whose vertex count $N_\mathrm{CL} \sim 4N^2$ goes from 31 to 188.
The value of $1/w$ is taken to be 8 or 20.
These parameters are discussed in App.~\ref{app:CL-embedding}, although their exact choice is not expected to change our results.

\begin{figure*}
    \includegraphics{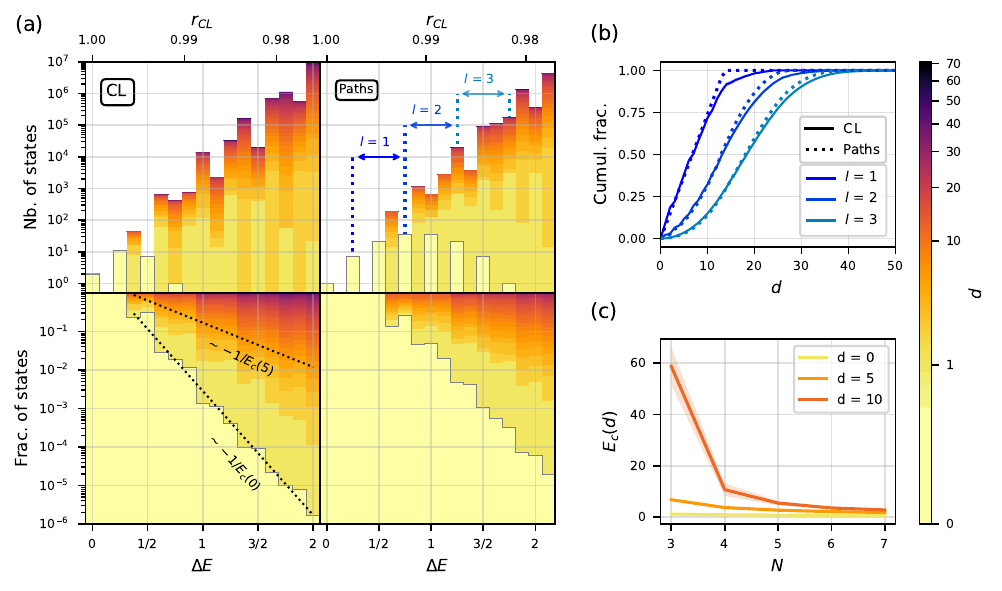}
    \caption{
    \textit{(a)} Low-energy sector \ac{DoS} \textit{(top)} and normalized cumulative \ac{DoS} \textit{(bottom)} for a random \ac{CL} graph with initial size $N = 7$ \textit{(left)} and for a path product $P_{4N}^N$ \textit{(right)}, both with weight bias $w = 1/8$. 
    Gray highlight: frontier of interpretable states. 
    Dotted line: exponential decrease $\sim \exp(-E/E_c(d))$ for $d \in \{0, 5\}$.
    \textit{(b)} 
    Cumulative normalized \ac{DoS} restricted to energy intervals with $\simeq l \in \{1, 2, 3\}$ domain walls and $w = 1/20$, both for the \ac{CL} and the path product.
    \textit{(c)} 
    Effect of graph size $N$ on the characteristic energy $E_c(d)$ at which the fraction of states with distance at most $d$ decreases for the \ac{CL}, with $w = 1/8$.
    Shaded areas indicate one standard deviation error bar.
    }
    \label{fig:CL-and-paths}
\end{figure*}

A typical example of low-energy \ac{DoS} is shown in Fig.~\ref{fig:CL-and-paths}~\textit{(a, top left panel)}, as a function of $\Delta E$ or equivalently of the approximation ratio $r$, with $1-r\propto \Delta E$.
The light yellow bars with dark contour on the left are the interpretable states, and they form a rescaled copy of the initial graph \ac{DoS}, with the optimal solution at the lowest energy and solutions of increasing approximation ratio as energy increases.
The other bars represent states with defects, which start appearing above the energy price of a domain wall $\Delta E = 1/2-w$.
Note that unless the initial graph is fully disconnected (a trivial case we can safely ignore), the initial MIS does not include all $N$ vertices; this implies the CL MWIS has at least one selected vertex with weight $1/2-w$ (encoding the deselection of a vertex of the initial graph).
Therefore, the lowest energy cost of a defect is $1/2-w$, instead of $1/2+w$ in path graphs.

In the \ac{CL}, the fraction of $d$-interpretable states $\tau_{d, \mathrm{CL}}(E)$ can be defined just as in Eq.~\eqref{eq:def-fraction-tau-path-graph}.
When computed in increasing energy windows, this fraction is shown in Fig.~\ref{fig:CL-and-paths}~\textit{(a, bottom left panel)}.
In good agreement with the analysis from Sec.~\ref{sec:path-embedding}, the low-energy \ac{DoS} can be separated in two regimes of energy.
Below the threshold $\Delta E = 1/2-w$, every state is interpretable.
Above the threshold, typical states quickly gain a high Hamming distance $d$.
We indeed observe that, for $d \ll \mathcal{O}(N^2)$, the fraction of $d$-interpretable states $\tau_d(E)$ decreases exponentially,
\begin{equation} \label{eq:exponential-decay-tau-d}
    \tau_{d, \mathrm{CL}}(E) \sim \exp(-\frac{\Delta E}{E_c(d)}),
\end{equation}
with a characteristic energy scale $E_c(d)$ above which states are unlikely to be at most $d$-interpretable.
This observation is strikingly similar to the behavior of path graphs studied in Fig.~\ref{fig:DoS-one-weighted-chain}.
As displayed in Fig.~\ref{fig:CL-and-paths} \textit{(c)}, $E_c(d)$ decreases toward zero with increasing problem size $N$.
This is observed with a remarkably low variability between random graph instances.
Similarly as for the path graph embedding, we conclude that states beyond the energy threshold $\Delta E = 1/2-w$ require some post-processing, making them harder to interpret.

We now argue that this pattern is independent from the implementation details of the \ac{CL}, and can pretend to some sort of universality.
More specifically, we show that the distribution of $d$ is reproduced by a simple gadget-less model of a product (disjoint unions) of path graphs.
We compare the \ac{CL} of initial size $N$ with the graph product $P_{4N}^{N}$ of $N$ weighted copies of $P_{4N}$, both graphs having the same $w$.
Using $N=7$ and $w = 1/8$, the \ac{DoS} of the path product shown in Fig.~\ref{fig:CL-and-paths} \textit{(a, top right panel)} is strikingly similar to the \ac{DoS} of the \ac{CL} \textit{(top left panel)}.
Next, we consider a small $w = 1/20$.
As in Fig.~\ref{fig:DoS-one-weighted-chain} \textit{(top right panel)}, in this regime, the \ac{DoS} features peaks located at half-integer values of $\Delta E$ (in units of $\delta$).
Neglecting some gadget-induced defects with energy $\simeq 3/2$, there is a correspondence between energy and the number $l$ of domain walls, i.e. $\Delta E \simeq l/2$ (see also App.~\ref{app:CL-embedding}).
The \ac{DoS} can thus be analyzed as a collection of peaks located near $\Delta E \simeq l/2$.
For $l \in \{1, 2, 3\}$, we plot in Fig.~\ref{fig:CL-and-paths} \textit{(b)} the normalized cumulative distribution of $d$ within each of these peaks.
With a low variability between the initial graphs (errors bars are smaller than the thickness of the line), we observe that the paths product model faithfully reproduces the interpretability features of the \ac{CL}.
This supports the idea that the combinatorial explosion of hard-to-interpret states is an intrinsic feature of the \ac{CL} scheme due to the introduction of ancilla nodes in the form of path graphs.
We expect this phenomenon to be somewhat universal and thus to exist in other embedding methods, such as graph subdivisions~\cite{pichler-2018-computational, dalyac-2023-exploring-graph-locality} (discussed in App.~\ref{app:subd-embedding}) and parity encoding~\cite{lechner-2015-LHZ, lanthaler-2023-LHZ}, where gadgets and paths graphs introduce extensive amounts of ancillae.

From this analysis, it is natural to ask whether the choice of weight profile $W_w$ may affect the exponential scaling Eq.~\eqref{eq:exponential-decay-tau-d}.
Indeed, the weight profile affects the distribution of defects and therefore their degeneracy. 
A study in this direction is presented in Appendix~\ref{app:mitigation}, suggesting modifications to the weight profile favor defects and should be avoided.

\section{Comparison to deselection strategy} \label{sec:deselection}

\begin{figure*}
    \includegraphics{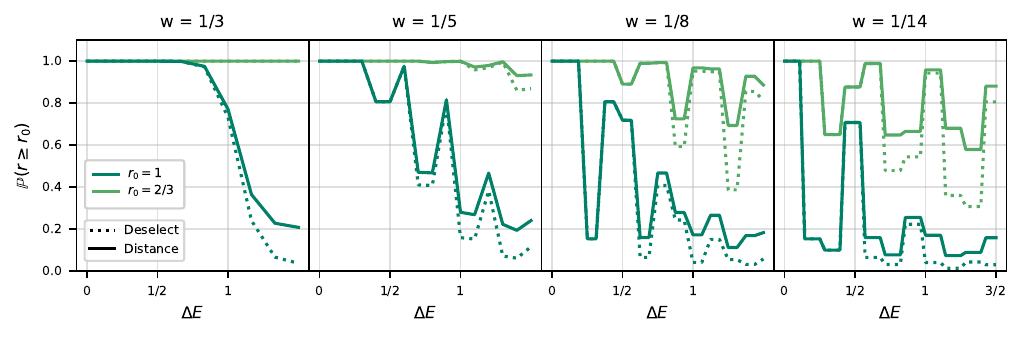}
    \caption{
    Probabilities of obtaining upon interpretation an approximation ratio $r \geq r_0$ for two post-processing strategies, with $r_0 \in \{ 2/3, 1\}$.
    Dotted lines: deselection strategy, for which each chain with defects is considered as unselected.
    Plain lines: interpretation of the nearest interpretable state with respect to the Hamming distance $d$.
    Data for one random graph with initial size $N = 7$ and \ac{MIS} size $\vert \text{MIS} \vert = 3$.
    }
    \label{fig:comparison-strategies}
\end{figure*}

We now compare the distance strategy with a second strategy, which is much cheaper to implement, but yields poorer quality results.
Recall that a \ac{CL} configuration can be seen as $N$ paths that are either in an antiferromagnetic ordering or have a domain-wall-like defect, as shown in Fig.~\ref{fig:MISs-even-path}~\textit{(a)}.
Furthermore, a path in an antiferromagnetic ordering can be considered as selected or unselected depending on which of its two endpoints is selected.
In that language, the \textit{deselection} strategy consists in finding all defective paths and declare them unselected.
Note that the deselection of chains is always possible without causing independence violations.
Indeed, the \ac{IS}s of the initial problem always form a stable set under the operation of deselecting any vertex.
Since the CL scheme is designed to respect the initial connectivity and embed all its possible ISs, the CL graph mirrors this property with chains instead of vertices.
As a result, the deselection strategy is guaranteed to return an interpretable \ac{CL} configuration as desired.

In terms of computational cost, finding all defects is easily done by walking through the \ac{CL} and spotting successions of unselected vertices which characterize defects (see Appendices~\ref{app:list-possible-defects} and~\ref{app:mitigation-results} for details).
Thus, the deselecting strategy costs $\order{N^2}$ to apply, so that we consider that it can be applied at any energy, unlike the distance strategy.

The drawback of this technique is that the output configurations are of lesser quality than the distance strategy.
To compare both strategies, we compute the approximation ratio $r$ obtained after interpretation in the \ac{CL} embedding of a random graph with $N = 7$ and \ac{MIS} size $\vert \text{MIS} \vert = 3$.
Then, for each technique and for two arbitrary thresholds $r_0 \in [0, 1]$, we show in Fig.~\ref{fig:comparison-strategies} the probability that a configuration taken at random within the energy window $[E_{\rm MWIS}, E_{\rm MWIS} + \Delta E]$ yields $r \geq r_0$.
In the case several nearest interpretable states exist, with possibly different values of $r$, our definition of the distance strategy becomes ambiguous. 
Our convention in this situation is to take the lowest approximation ratio, so that our results still give a lower bound of the advantage of the distance strategy over the deselection strategy.
We also checked these cases are rare (less than 2.2\% of states displayed in Fig.~\ref{fig:comparison-strategies}), which legitimates our conclusions irrespective of the chosen convention.

When $\Delta E \geq 1/2$, we observe in Fig.~\ref{fig:comparison-strategies} that the distance strategy tends to output better approximation ratios than the deselection strategy.
This advantage is expected and we explain it as follows.
First, at least in the low-energy spectrum under study, the distance strategy almost always leaves non-defective chains untouched.
We checked numerically that CL vertices located in non-defective paths had to be flipped in only 0.8\% of the configurations from Fig.~\ref{fig:comparison-strategies}.
Consequently, for the vast majority of cases, the distance strategy outputs configurations of the initial graph with at least as many selected vertices than the deselecting strategy, which translates into better approximation ratios for the distance strategy.

It is striking that the difference in approximation ratios for the two strategies is not very large (see also Fig.~\ref{fig:DoS-both-strategies} in App.~\ref{app:approx-ratio}), given the huge difference in computational complexity.
We explain such good results for the deselection strategy as follows.
Let us consider an \ac{MWIS} state in the \ac{CL}, some of its chains being selected and the other being unselected.
If then a domain wall is added to make a single chain defective, both endpoints of the chain become selected, as pictured in Fig.~\ref{fig:MISs-even-path}.
However, these two endpoints have a different weight $1/2 \pm w$ due to Eq.~\eqref{eq:weight-profile-chain}.
Thus, one can see the added defect typically increases the configuration energy by an amount $1/2 \pm w$, where the $\pm w$ term depends solely on whether the chain was previously selected ($+w)$ or not ($-w$). 
As a result, we expect defects will energetically favor unselected chains.
When using the deselection strategy, such defects are \textit{harmless}: adding one does not change the approximation ratio after interpretation, since the chain it is applied to was by definition unselected.
This is not the case of \textit{harmful} defects, i.e. those added on selected chains.
Since they incur a higher energy cost $1/2+w$, they are energetically penalized when compared to harmless defects, a mechanism which explains to some extent the success of the deselection strategy.

The previous argument is valid for any $w > 0$, even for $w > 1/2$, a non-standard choice of parameter which may be beneficial and which we discuss in Appendix~\ref{app:approx-ratio}.
Nonetheless, we continue using $w < 1/2$ in the rest of this article.

\paragraph{Lower bound for $r$ with the deselection strategy.}
We can find a lower bound for the approximation ratio obtained through the deselection strategy.
Consider a configuration $S_\mathrm{CL}$ of the \ac{CL}.
Through the deselection strategy, its defective chains are replaced by unselected chains, defining an interpretable \ac{CL} configuration $S_\mathrm{des}$.
Since defects cost energy, deselecting defective chains lowers the energy: $E(S_\mathrm{des}) \leq E(S_\mathrm{CL})$.

By construction, the \ac{CL} weight profile ensures that flipping any  chain from selected to unselected (which is always possible without breaking the independence constraint) increases the energy by $2w$. 
In particular, starting from the \ac{MWIS} with its $\vert \text{MIS} \vert$ selected chains, any interpretable configuration with $N_\mathrm{sel}$ selected chains has $\vert \text{MIS} \vert - N_\mathrm{sel}$ less selected chains than the \ac{MWIS} and thus has an energy $E_\mathrm{MWIS} + 2w \cdot (\vert \text{MIS} \vert - N_\mathrm{sel})$.
Here we recall that $\vert \text{MIS} \vert$ is the  \ac{MIS} size of the initial graph. Since the approximation ratio upon interpretation is $r  = N_\mathrm{sel}/\vert \text{MIS} \vert$, the energy of any interpretable configuration is linearly related to its approximation ratio upon interpretation.
This applies in particular to $S_\mathrm{des}$: 
\begin{equation}
    r
    =
    1 - {
        E(S_\mathrm{des}) - E_\mathrm{MWIS}
        \over
        2w \cdot \vert \text{MIS} \vert
    },
\end{equation}
where we assume $w < 1/2$ as always.
Finally, using $E(S_\mathrm{des}) \leq E(S_\mathrm{CL})$ we deduce a lower bound on the approximation ratio provided by the deselection strategy:
\begin{equation}
    \label{eq:bound-approx-ratio-deselecting}
    r \ge 1 - \frac{E(S_\mathrm{CL}) - E_\mathrm{MWIS}}{2w\cdot \vert \text{MIS} \vert}.
\end{equation}
We point out that this bound cannot be improved since it is saturated whenever $S_\mathrm{CL}$ is interpretable.

\section{Discussion: asymptotic behavior} \label{sec:discussion} 

For larger graphs, the previous methodology becomes intractable due to the combinatorial explosion of sub-optimal solutions.
To circumvent this, we use analytical argument to extend some keys observations to the asymptotic limit $N \longrightarrow + \infty$.
In this regime we expect the distance strategy to be hard to apply, as a result of App.~\ref{app:algo_NIS}, so we focus on the deselection strategy.
With this in mind, we link the previous observations to complexity requirements on a quantum \ac{UD}-\ac{MWIS} solver in the case of the deselection strategy.
As a first step, we prove that using the \ac{CL} embedding scheme leads to a degradation of the approximation ratio, namely to a magnified error upon interpretation.

\subsection{Error amplification}

From the previous section, a key quantity here is the energy $E(S_\mathrm{CL})$ of a sampled \ac{CL} configuration obtained by approximate \ac{UD}-\ac{MWIS} optimization.
Let us assume that the quantum solver produces a number of defects (such as domain walls) that scales as $N^\gamma$ for some constant $\gamma \geq 0$.
For instance, under the Kibble-Zurek mechanism~\cite{ebadi-2022, kibble-1976-cosmo, zurek-1985-cosmological, zurek-1996-cosmological, bunkov-2000-topological-defects}, we would get a finite density of defects, hence $\gamma = 2$.
Since the energy cost of each defect is $\Theta(1)$, this implies we get configurations of energy
\begin{equation}
E
\; = \; 
E_\mathrm{MWIS} + \Theta(N^\gamma)
\label{eq:scaling}
\end{equation}
Using Eq.~\eqref{eq:bound-approx-ratio-deselecting}-\eqref{eq:scaling}, the deselection strategy leads to an approximation ratio lower bound
\begin{equation}
\label{eq:result-r-gamma}
    r \ge 1 - \frac{\Theta(N^\gamma)}{2 w |\text{MIS}|} = 1 - \Theta(N^{\gamma-1}),
\end{equation}
where we used, in the equality, the standard assumption $\vert \text{MIS} \vert = \Theta(N)$, which is verified in typical families of graphs of interest, such as bounded-degree graphs and some families of random graphs~\cite{caro-1979-wei-theorem, wei-1981-caro-theorem, coja-2011}.

We distinguish between two regimes.
When $\gamma < 1$, the deselection strategy ensures an approximation ratio asymptotically close to 1.
When $\gamma > 1$, as expected in the current state of the art, the strategy asymptotically fails: the number of defects becomes infinitely larger than the number of chains, such that almost certainly every chain gets a defect and is deselected upon interpretation, yielding $r = 0$.

In the \ac{CL}, let us denote by $\vert \text{MWIS}_\mathrm{CL} \vert$ the \ac{MWIS} total weight.
The approximation ratio $r_\mathrm{CL}$ of an \ac{IS} in the \ac{CL} is related to its energy through
\begin{equation}
    r_\mathrm{CL} = 1 - {E - E_\mathrm{MWIS}
    \over \vert \text{MWIS}_\mathrm{CL} \vert}.
\end{equation}
With the deselection strategy, the approximation ratio after interpretation $r$, and before interpretation $r_\mathrm{CL}$, are thus related through Eq.~\eqref{eq:bound-approx-ratio-deselecting} by
\begin{equation} \label{eq:scaling-r-rCL}
    1-r
    \le
    {
    (1 - r_\mathrm{CL}) \vert \text{MWIS}_\mathrm{CL} \vert
        \over
    2w \cdot \vert \text{MIS} \vert
    }
    =
    \Theta\left( N (1-r_\mathrm{CL}) \right),
\end{equation}
where we used the fact that $\vert \text{MWIS}_\mathrm{CL} \vert = \Theta(N^2)$.
Importantly, Eq.~\eqref{eq:scaling-r-rCL} shows that solving the \ac{MWIS} problem means solving the embedded \ac{MIS} problem with an error up to $N$ times higher (within a bounded factor).
This amplification of the error exemplifies the limitation of embedding schemes caused by the interpretation step.

\subsection{Approximability theory consequences}

We now discuss the consequences of the error amplification in Eq.~\eqref{eq:scaling-r-rCL} due to the deselection strategy in terms of approximability theory.
This theory classifies the hardness to find approximate solutions, namely the hardness of getting close to $r=1$.
We will focus on three classes of approximation: the \ac{fPTAS} class, which describes problems for which an approximation algorithm exists that guarantees an error at most $\epsilon$ in run time polynomial in \textit{both} the instance size and $1/\epsilon$; the \ac{PTAS} class, which also ensures $r\geq 1-\epsilon$ but with a run time polynomial in the instance size only, and that can be worse than polynomial in $1/\epsilon$; and the APX class, which only achieves $r \geq c$ for some constant $c < 1$.
Note that the terms \ac{PTAS} and \ac{fPTAS} also denote the algorithms themselves (in addition to the approximation classes). 

In terms of classical approximation algorithms, the \ac{UD}-\ac{MWIS} problem belongs to the \ac{PTAS} class~\cite{Erlebach-PTAS-UD-MWIS, matsui-1998-approximating-MIS-on-UDG, Das-2015-approximating-MIS-on-UDG, Nandy-2017-approximating-MIS-on-UDG, bonamy-2018-eptas-UBMIS}, while the general MIS problem is APX-complete~\cite{Hastad-1999-hardness-approximating-MIS, garey-1978-MIS-strongly-NP-hard, bazgan-2005-approximating-MIS-is-hard}.
\ac{MIS} is thus believed to admit no \ac{PTAS}. 
In fact, a corollary to the PCP theorem asserts that the MIS problem cannot be in the \ac{PTAS} class (unless P=NP).
Likewise, according to the quantum PCP conjecture~\cite{aharonov-2013-quantum-pcp}, MIS should remain in the quantum equivalent of the APX class even with quantum algorithms.

Let us examine how these general statements relate to our findings.
Classical \acp{PTAS} exist for \ac{UD}-\ac{MWIS}, and due to the PCP theorem and quantum PCP conjecture, there cannot be any classical, and likely any quantum, \ac{PTAS} for MIS.
This means that either some degradation in the approximation ratio must happen during the interpretation, or the interpretation algorithm (or the embedding itself) is not polynomial in the problem size. 
The distance strategy falls in the second category (it is exponential), while the deselection strategy falls in the first category, as it is polynomial in the problem size.

For the deselection strategy, this is embodied by Eq.~\ref{eq:scaling-r-rCL}, showing that an error $\epsilon_\mathrm{CL} = \epsilon/N$ on the \ac{UD}-\ac{MWIS} resolution translates to an error up to $\epsilon$ on the \ac{MIS} problem.
If an \ac{fPTAS} were to exist for \ac{CL} graphs, it would give an \ac{fPTAS} for \ac{MIS}, because $\Poly(N_\mathrm{CL},1/\epsilon_\mathrm{CL}) = \Poly(N, N/\epsilon)$. 
However, this is unlikely due to the earlier PCP considerations, excluding the existence of a quantum or classical \ac{fPTAS} for CL, and so for \ac{UD}-\ac{MWIS} as a whole.
On the other hand, a \ac{PTAS} for CL, which is not an \ac{fPTAS}, would only provide a super-polynomial run time, e.g. $\exp(1/\epsilon_\mathrm{CL}) = \exp(N/\epsilon)$, and therefore would not give a \ac{PTAS} for \ac{MIS}, in agreement with the PCP theorem.
To sum up, PCP considerations rule out the existence of a PTAS for MIS and of an fPTAS for CL graphs, and so for UD-MWIS in general.

This discussion has even more practical consequences, i.e. in terms of number of defects.
Going back to the assumption of Eq.~\eqref{eq:scaling}, $\gamma$ is related to the approximation ratio through Eq.~\eqref{eq:result-r-gamma}.
Consider then an hypothetical quantum solver for the CL, running polynomially in $N$ and with a number of defects scaling as $\Theta(N^\gamma)$.
With the deselection strategy, it would solve MIS with an error $\epsilon = 1 - r \le \Theta(N^{\gamma-1})$.
If $\gamma \geq 1$, like in the standard Kibble-Zurek scenario where $\gamma = 2$, the explosion in the number of defects leads to approximation ratios so poor that no error $\epsilon < 1$ can be guaranteed.
If $\gamma < 1$, however, this algorithm would induce a PTAS for MIS, because the error can be made arbitrarily low by duplicating the initial MIS graph.
Indeed, having $M$ copies of the initial graph leads to a CL with $M\cdot N$ chains, guaranteeing an error $\le \Theta((MN)^{\gamma-1})$ that vanishes at large $M$, but still solved in $\Poly(MN) = \Poly(N)$ time. 
Since this is ruled out by the previous discussion, we conclude that, very likely, there exists no quantum or classical CL solver running polynomially in $N$ with a number of defects scaling as $o(N)$.
This statement naturally extends to UD-MWIS in general.

In conclusion, no improvement in approximability class is likely to happen by using an embedding of MIS in a larger \ac{UD}-\ac{MWIS} problem.
This represents a very practical embodiment of the PCP theorem and its quantum equivalent for embedding strategies.
This of course does not prevent \emph{heuristic} algorithms, namely algorithms with no run time guarantees, from achieving good enough approximation ratios.

\section{Conclusion and outlook} \label{sec:conclusion}

In this work, we studied quantitatively the robustness to errors of embedding techniques to solve the \ac{MIS} problem on Rydberg atom machines.
We focused on the \ac{CL} construction as a case study.
We highlighted the difficulty of interpreting, or correcting, defective solutions to the \ac{CL}---which unavoidably results from realistic adiabatic procedures---into solutions to the original \ac{MIS} problem, while retaining a high approximation ratio.

We studied two opposite strategies.
The \emph{distance} strategy produces high quality solutions in the presence of few defects, but is untractable with large number of defects.
The \emph{deselection} strategy, on the other hand, is straightforward to apply with any number of defects, at the expense of a worst approximation ratio.
Nevertheless, asymptotically at large $N$, these strategies require an extremely small number of defects scaling as $\Theta(1)$ and $o(N)$ respectively, to be tractable and produce non-zero approximation ratios. 
With the deselection strategy, this is a consequence of the error amplification induced by the embedding, which is linked to an enlarged Hilbert space with $\Theta(N^2)$ qubits.

Our observations on the \ac{CL} graphs seem independent from the connectivity between chains, suggesting some form of universality, when approximating embedding graphs.
As a consequence, we expect similar results when extending our approach to integer factorization~\cite{park-2024-factorization-with-rydberg, nguyen-2023}, \ac{QUBO} problems~\cite{nguyen-2023, byun-2024-QUBO-with-Rydberg}, and related embedding techniques~\cite{pichler-2018-computational, pichler-2018-experimental, ebadi-2022, dalyac-2023-exploring-graph-locality, kim-2022-3d-rydberg-wires, lechner-2015-LHZ, lanthaler-2023-LHZ}.
We also expect the $o(1/N)$ requirement in the density of defects to apply to other strategies which output better approximation ratios than the deselection strategy.

Using approximability theory, we demonstrated that getting a $o(1/N)$ density of defects in polynomial time is unlikely to be feasible, due to the quantum PCP conjecture.
Otherwise, the combination with e.g. the deselection strategy would produce an fPTAS for MIS, which is ruled out by this conjecture.
Further work would be required to study more precisely how this limitation is embodied in adiabatic algorithms, in particular in terms of energy gaps.
A more realistic goal could thus be to outperform existing approximate classical \ac{PTAS} or heuristics for \ac{UD}-\ac{MWIS}.

\acknowledgments

We thank  F. Meneses, A. Oliveira and H. Pichler for useful exchanges.
This work was supported by the European High-Performance Computing Joint Undertaking (JU) under grant agreement No 101018180 (HPC-QS).
This publication has received funding under Horizon Europe programme HORIZON-CL4-2022-QUANTUM-02-SGA via the project 101113690 (PASQuanS2.1).


%

\newpage
\appendix 
\onecolumngrid

\section{\ac{DoS} in paths graphs} \label{app:DoS-paths-graphs}

This section introduces tools from graph theory that provide a concise notation for the \ac{DoS} in \ac{MWIS} problems. 
In App.~\ref{ap:DoS_chains}, this is used to characterize the \ac{DoS} for path graphs.
The following sections expand on this formalism to treat the case of weighted paths and \ac{CL} graphs.

\subsection{Independence polynomials}

To study the non-interpretable states that arise from suboptimal \ac{UD}-\ac{MWIS} optimization, it is convenient to borrow several notions from the fields of combinatorics and graph theory.
The \textit{independence polynomial} $\IP{G, W; \X}$ is a generalized polynomial, whose coefficient at any (possibly non-integer) degree $w \in \mathbb{R}_+$ is the number of \ac{IS}s whose vertices have a total weight $w$:
\begin{equation} \label{eq:def-IP-weighted-graph}
    \IP{G, W; \X} 
    = 
    \sum_{S \in IS(G)}\X^{\sum_{i \in S}W(i)} 
    = 
    \text{Tr}\left(\X^{-\mathcal{H}_{\text{MWIS}}}\right). 
\end{equation}
As a reminder, the energy scale $\delta = 1$ is omitted.
We will frequently drop the notation of the weight profile $W$ in $\IP{G, W}$ if the graph is unweighted, as well as the notation of the indeterminate $\X$.
One can relate the independence polynomial with the canonical partition function $\mathcal{Z}(\beta)$, introducing a dimensioning factor $\beta$ which represents the thermodynamical inverse temperature:
\begin{equation} \label{eq:equiv-partition-function-IP}
        \mathcal{Z}(\beta) 
        \; = \;
        \text{Tr}\left( e^{-\beta \mathcal{H}_{\text{MWIS}}}\right) 
        \; = \; 
        \IP{G, \, W ; \; \X = e^{-\beta}}.
\end{equation}

Conversely, the independence polynomial characterizes the \ac{DoS} $\rho$ seen as a distribution.

The partition function depends on the \ac{DoS} as
\begin{equation}
    \mathcal{Z}(\beta) = \int \d E e^{-\beta E} \rho(E).
\end{equation}
Inverting this relation and replacing $\mathcal{Z}(\beta)$ by $\IP{z}$ with the change of variable $z = e^{-\beta}$, we get
\begin{equation}
    \rho(E) 
    \; = \;
    \int_{-i\infty}^{+i\infty} \frac{\d \beta}{2i\pi} \mathcal{Z}(\beta) e^{\beta E}
    \; = \;
    \frac{1}{2i\pi} \oint_{\mathcal{C}}\IP{z}z^{E - 1} \d z.
\end{equation}
where $\mathcal{C}$ is an anticlockwise contour in the complex plane around the origin.

Both formulations being equivalent, we choose to work with independence polynomials due to the simplicity of notations.

\subsection{Unweighted paths} \label{ap:DoS_chains}

Let $P_{N}$ be the path graph with $N \in \mathbb{N}$ vertices.
A previous result from Refs~\cite{arocha-1984-propriedades, Levit-Mandrescu-2003} is that
\begin{equation} \label{eq:ip-path-polynomial-N}
    \forall N \geq 0, \quad \IP{P_N} = \sum_{k \geq 0}{N - k + 1 \choose k}\X^k,
\end{equation}
where the summand is zero whenever $k > \lfloor (N+1)/2 \rfloor$.
In particular the two first path polynomials are obtained from a combinatorial enumeration:
\begin{equation}
    \IP{P_0} = 1, \quad \IP{P_1} = 1 + \X,
\end{equation}
which is quickly verified by a combinatorial enumeration of the \ac{IS}s in $P_0$ and $P_1$.
For the sake of completeness, we mention that Eq.~\eqref{eq:ip-path-polynomial-N} can be proved by induction with a well-known \textit{Fibonacci recurrence relation} from Ref.~\cite{arocha-1984-propriedades}:
\begin{equation} \label{eq:rec-relation-path-IP}
    \forall N \geq 1, \quad \IP{P_{N+1}} = \IP{P_{N}} + \X\cdot\IP{P_{N-1}}.
\end{equation}
Moreover, we provide an apparently new derivation of Eq.~\eqref{eq:ip-path-polynomial-N} based on combinatorics.
Given $N$, $k$, an \ac{IS} in $P_N$ with $k$ selected vertices can be built by choosing how many unselected vertices are on the right and left of each selected vertex. 
This amounts to choosing two non-negative integers $l, r$ and $(k-1)$ positive integers $m_1, \dots, m_{k-1}$ such that
\begin{equation}
    l + m_1 + \dots + m_{k-1} + r = N-k.
\end{equation}
Here $(l, m_1, \dots, m_{k-1}, r)$ characterizes the \ac{IS} in $P_N$ made by the succession of $l \ge 0$ unselected vertices, then one selected vertex, then $m_1 \ge 1$ unselected vertices (ensuring the independence constraint), and so on.
This clearly defines a one-to-one correspondence between the elements of $\IS{P_N}$ and the possible sequences $(l, m_1, \dots, m_{k-1}, r)$.
By application of the so-called \textit{stars and bars theorem} from combinatorics, one retrieves the number $\left[\X^k\right]\IP{P_N}$  of possible choices for $(l, m_1, \dots, m_{k-1}, r)$ with the same result as in Eq.~\eqref{eq:ip-path-polynomial-N}.

\subsection{Weighted paths}

Let $w \in (0, 1/2)$ be a weight bias.
We endow the path $P_{2N}$ with the weight profile $W_w$ as defined in Eq.~\eqref{eq:weight-profile-chain}.
By disjunction over which vertices in $\{v_1, v_{2N}\}$ are selected, all \ac{IS}s can be partitioned in four categories.
For instance, if  $v_1$ is selected and $v_{2N}$ is not, then this lets $2N-3$ remaining vertices that can take any independent configuration from the path $P_{2N-3}$.
Since $v_1$ contributes to a weight $1/2+w$, the polynomial $\X^{{1 \over 2} + w}\IP{P_{2N-3}}$ characterizes the \ac{DoS} of this category of \ac{IS}s.
The three other categories can be treated likewise to obtain the independence polynomial for the weighted path $P_{2N}$:

\begin{equation}\label{eq:IP-chain-weighted}
    \IP{P_{2N}, W_{w}} 
    \; = \;
    \IP{P_{2N-2}}
    +
    \X^{{1 \over 2}}\left( \X^w + \X^{-w}\right)\IP{P_{2N-3}}
    +
    \X\IP{P_{2N-4}}.
\end{equation}

Again, this is another way to write the \ac{DoS} of the corresponding Hamiltonian $\mathcal{H}_{\text{MWIS}}$.
The highest-degree terms count in this regard $\sim N$ non-interpretable configurations located at an energy of only $1/2-w$.
This proves that the manifold of lowly-excited \ac{IS}s in $P_{2N}$ is filled with non-interpretable configurations.

\section{Crossing lattice} \label{app:CL-embedding}

This section is devoted to some formal aspects of the \ac{CL} that are discussed in our main text. 
First, we introduce several definitions and  enumerate a family of states in the \ac{CL}, which serves as a lower bound for its \ac{DoS}.
We conclude by a discussion of the numerical parameters used in our study.

\subsection{Notations}

\begin{figure}
    \centering
    \includegraphics[width=1.0\linewidth]{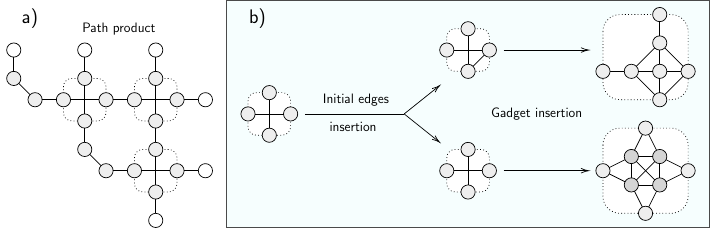}
    \caption{
    \textit{a)} The non-planar \ac{CL} is a weighted path product $P_{2N+2}^N$. \textit{b)} Upon insertion of new edges, the resulting \textit{non-planar \ac{CL} embedding graph} satisfies the criteria for an embedding graph except for the planarity. Upon insertion of suitable gadgets, one retrieves a \ac{CL} embedding graph.
    }
    \label{fig:CL-construction}
\end{figure}

Let $G$ be an unweighted graph with $N \gg 1$ vertices.
We consider the \ac{CL} embedding of $G$ for an arbitrary ordering of the vertices in $G$.
Our goal is not to formalize its construction, but to introduce the vocabulary required in our derivations.
Given a path $P_{2N} = \{v_1, \dots, v_{N}\}$ with edges between $v_i$ and $v_{i+1}$, the vertices $v_1, v_3, \dots v_{2N-1}$ are said to be \textit{odd vertices} while $v_2, v_4, \dots, v_{2N}$ are called \textit{even vertices}.
We use the ordering convention that if $P_{2N}$ is weighted by $W_w$, then $v_1$ is always the vertex with weight $1/2+w$, in accordance with the main text.

Let us consider the path product $P_{2N+2}^N$, in which each path has the weight profile $W_w$ for some $w$ and is called a \textit{chain}.
As shown in Fig.~\ref{fig:CL-construction} for the case $N = 3$, this graph can be drawn in 2D by arranging the $N$ paths along the lines of a grid lattice, following the construction of Ref.~\cite{nguyen-2023}.
Instead of introducing \ac{UD} gadgets to replicate the edge structure from $G$, one can directly insert copies of the edges in $G$.
We call the resulting graph a \textit{non-planar \ac{CL}} embedding graph of $G$ and denote it by $G_{np}$.
By identification of $P_{2N+2}^N$ as a subgraph of $G_{np}$, the $N$ chains are still defined in $G_{np}$.
Clearly, this construction of $G_{np}$ satisfies all the requirements for an embedding with the usual interpretable states defined by having each chain in an antiferromagnetic state with $N+1$ selected vertices.

Given some configuration $S_{np} \in \IS{G_{np}}$, we denote by $\mathcal{C}_i$ the subset of vertices in $S_{np}$ that belong to the $i$-th chain in $G_{np}$.
Here the ordering of the chains is the same ordering than for the vertices $\{v_i, 1\leq i \leq N\}$ of $G$ they embed.
Given some $\mathcal{C}_i$ in a chain that embeds a vertex $v_i \in G$, we say that this chain is \textit{selected} if $\mathcal{C}_i$ is a subset of the \ac{MWIS} in $G_{np}$, namely if the only selected vertices are odd.
Because the chain itself is a copy of $P_{2N+2}$ with weight profile $W_w$, this \ac{MWIS} is indeed unique.
Conversely, if $\mathcal{C}_i$ only has even selected vertices, then it is said to be \textit{unselected}.
Finally, we introduce two independence polynomials
\begin{equation} \label{eq:def-polynomials-a-pm}
    A_{N, \pm} = \left(1 + \X^{\frac12\pm w} \right)\left( 1 + \X \right)^{N}.
\end{equation}
The combinatorial interpretation of these polynomials is that $A_{N, +}$ ($A_{N,-}$) describes all possible subsets of the (un)selected configuration in a weighted chain.
Note that except for the empty chain which, by definition, is both selected and unselected, this enumerates all configuration where all vertices have the same parity.

\subsection{Lower bound on the \ac{DoS} of \ac{CL} graphs}

Due to the dependence of $G_{np}$ in $G$, and to the hardness of computing the independence polynomials for grid graphs, we are only interested in approximating $\IP{G_{np}}$.
This can be done by enumerating as many \ac{IS}s in $G_{np}$ with their respective weight, while ensuring no \ac{IS} is counted twice and each \ac{IS} is, indeed, an independent set.
Intuitively, the family of \ac{IS}s that we will build in this section is made by considering interpretable states and allowing in each chain different defects, depending on wether the chain embeds an initial vertex that is selected.
This process will ensure that all built configurations remain independent, in spite of the gadgets connecting the chains.

Let $S \in \IS{G}$, and assume $v_i \neq v_j$ are two vertices in $S$.
If $v_i$ and $v_j$ are not adjacent in $G$, then there is in $G_{np}$ no edge between the chain that embeds $v_i$ and the chain that embeds $v_j$.
Thus both chains can be in any of the \ac{IS}s of $P_{2N+2}$, without violating the independence criterion.
Conversely, let us assume that $v_i$ and $v_j$ are adjacent in $G$.
Then one of the odd vertices in $\mathcal{C}_i$ is adjacent in $G_{np}$ to one of the odd vertices in $\mathcal{C}_j$.
In that case if $\mathcal{C}_i$ is in a configuration where only even vertices are selected, and $\mathcal{C}_j$ itself is in an \ac{IS} of $P_{2N+2}$, then there is no independence violation.
This can be generalized to the $N$ vertices of $G$ as follows.
Let $k = \vert S\vert$ be the number of selected vertices, and $i_1, \dots, i_k$ the indices of the vertices in $G$ that are in $S$.
By our previous argument, if each chain $\mathcal{C}_{i_1}, \dots, \mathcal{C}_{i_k}$ is in an \ac{IS} of $P_{2N+2}$, and every other chain is unselected, then this defines an \ac{IS} $S_{np}$ in $G_{np}$.
For a given $S$, all the \ac{IS}s that are built by this process define a \ac{DoS} whose corresponding independence polynomial is 
$\IP{P_{2N+2}, W_w}^kA_{N, -}^{N-k}$. 

As stated above, this \ac{DoS} depends on one specific $S \in \IS{G}$.
We are now interested in obtaining a \ac{DoS} which corresponds to all the possible $S \in \IS{G}$, and computing its independence polynomial.
This is done by adding all the contributions from each possible $S$, with the only precaution that each configuration can be enumerated at most once.
To enforce this condition, let us consider that if $v_i \in S$,  then $\mathcal{C}_i$ can take any configuration but cannot be unselected.
Intuitively, this allows to invert our construction process by retrieving $S \in \IS{G}$ from the chains that have at least one odd selected vertex.
By summation over all possible $S \in \IS{G}$, our enumeration describes a family of \ac{IS}s in $G_{np}$ whose \ac{DoS} is represented by the polynomial
\begin{equation}
    \IP{G_{np}}_{lb} 
    = 
    \sum_{S \in \IS{G}}
    \left(\IP{P_{2N+2}, W_w} - A_{N, -}\right)^{\vert S \vert}A_{N, -}^{N-\vert S \vert}
    =
    A_{N, -}^N
    \IP{G ; \, 
    \X = {\IP{P_{2N+2}, W_w}\over A_{N, -}} - 1},
\end{equation}
where the subscript $lb$ stands for \textit{lower bound}, in the following sense:
\begin{equation}
    \forall w \in \mathbb{R}_+, \quad
    \left[\X^w\right]\IP{G_{np}}_{lb} 
    \; \leq \; 
    \left[\X^w\right]\IP{G_{np}}.
\end{equation}

The previous lower bound can be generalized to the \ac{CL}.
As pictured in Fig.~\ref{fig:CL-construction}, the \ac{CL} is equivalent to the non-planar \ac{CL} upon insertion of two types of gadgets, so it remains to lower bound the \ac{DoS} of possible configurations for each gadget given the configuration of the two chains around it.
In both types of gadgets, this leaves the possibility to select at most one vertex in the central complete subgraph in the case of interpretable states, with weight 1 (crossing-with-edge) or 2 (crossing).
If $\vert E\vert$ is the number of edges in the initial graph $G$, then this yields a lower bound $\IP{G_\mathrm{CL}}_{lb}$ for the \ac{DoS} in the \ac{CL} as follows:
\begin{equation} \label{eq:IP-CL-lower-bound}
    \IP{G_\mathrm{CL}}_{lb} 
    =
    (1+\X)^{\vert E\vert}
    \times
    \left( 1+\X^2\right)^{N(N-1)/2-\vert E\vert
    }
    \times
    \IP{G_{np}}_{lb},
\end{equation}
with as before
\begin{equation}
    \forall w \in \mathbb{R}_+, \quad
    \left[\X^w\right]\IP{G_\mathrm{CL}}_{lb} 
    \; \leq \; 
    \left[\X^w\right]\IP{G_\mathrm{CL}}.
\end{equation}

More gadget-induced defects can be taken into account by considering separately the cases where zero, one or two boundary vertices of a gadget are selected.
The opposite stance can also be taken by first studying the product of the $N(N-1)/2$ gadgets, and then retrieving the \ac{CL} by inserting small path subgraphs.
However these approaches are less tractable and typically require more knowledge of $G$ like its adjacency matrix.

\subsection{Approximation ratio} \label{app:approx-ratio}

In the scope of the \ac{CL} embedding, the relevant approximation ratio is the one obtained after interpretation, $r$.
For an interpretable configuration $S_\mathrm{CL}$, it is always $r(f^{-1}(S_\mathrm{CL}))$.
For non-interpretable configurations, it depends on the interpretation strategy. Using the distance strategy, it would be the approximation ratio of the nearest interpretable configurations.
In the (rare) case of several equidistant configurations, we chose the approximation ratio of the last one with respect to the order returned by the solver.
Note that Sec.~\ref{sec:deselection} uses a different convention (choosing the smallest approximation ratio), and justifies that the exact convention should not change the results.

\begin{figure}
    \centering
    \includegraphics
    {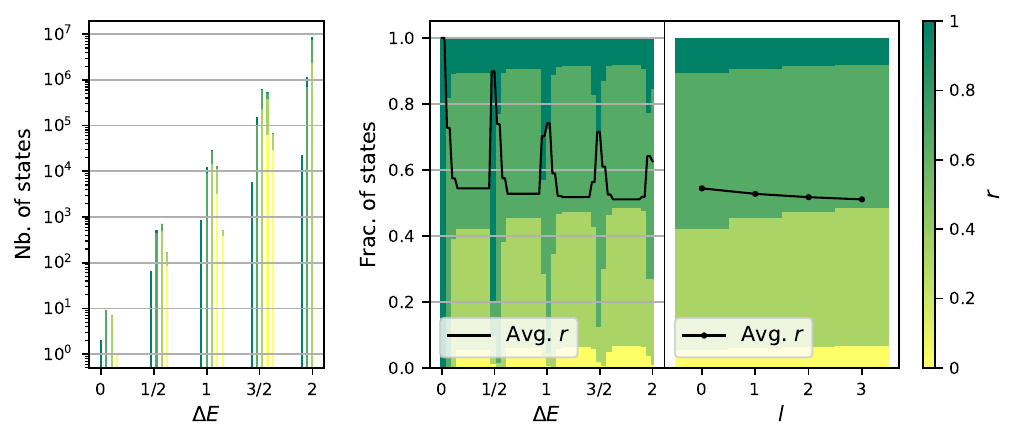}
    \caption{
    approximation ratio $r$ after interpretation in a random CL with \ac{CL} graph with initial size $N = 7$ and weight bias $w = 1/20$.
    Here the approximation ratio is obtained with the distance strategy.
    \textit{(Left)}
    Low-energy \ac{DoS}.
    \textit{(Middle)}
    Repartition of $r$ in increasing energy windows $[E_{\text{MWIS}}, E]$ and for the same parameters.
    \textit{(Right)}
    Repartition of $r$ in energy windows defined by $\vert \Delta E - l/2\vert \leq 1/4$ for $l = 0, \dots, 3$.
    }
    \label{fig:approximation-ratio}
\end{figure}

In Fig.~\ref{fig:approximation-ratio} \textit{(left panel)}, we plot the low-energy \ac{DoS} obtained with the same \ac{CL} graph than in Fig.~\ref{fig:CL-and-paths} \textit{(a, left column)}, with $w = 1/20$.
Here the color indicates the approximation ratio $r$ produced after applying the distance strategy.
In this regime where $w \ll 1$, we observe that the approximation ratio quickly decreases towards a value which is the average $r$ over all \acp{IS} of the initial graph, with sharp increases at energies allowing new defects.
This is explained by the fact that the \ac{CL} \ac{DoS} embeds a copy of the initial \ac{DoS} at energies arbitrarily close to the ground-state energy when $w \longrightarrow 0^+$.
The general trend observed as a function of the typical number $l$ of domain walls is shown in Fig.~\ref{fig:approximation-ratio} \textit{(middle and right panel)}.
It indicates a slight decrease of the approximation ratio when $l$ increases, and this behavior is consistent across all other random graphs (not shown).
This raises the question of whether different gadgets would lead to an increase of the variation of $r$ with respect to $l$, a question we leave for further work.\\

\begin{figure}
    \centering
    \includegraphics[width=1.0\linewidth]{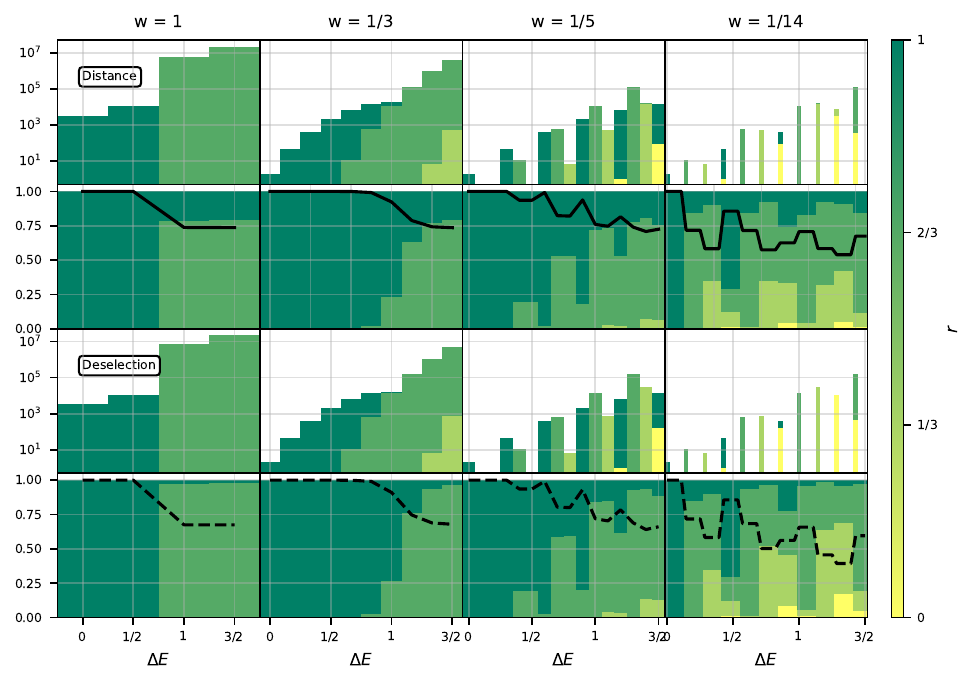}
    \caption{
    Distribution of the approximation ratio obtained with the distance strategy \textit{(first two rows)} and with the deselection strategy \textit{(last two rows)} as described in Sec.~\ref{sec:deselection} and with the same graph used for Fig.~\ref{fig:comparison-strategies}.
    As before, the distribution for each value of $1/w \in \{ 1, 3, 5, 14\}$ is shown with the actual \ac{DoS} (first and third rows) and after cumulating the \ac{DoS} in increasing energy windows $[E_\mathrm{MWIS}, E]$ (second and last rows).
    Black lines: average $r$ obtained when sampling at random a configuration in the increasing energy window.
    }
    \label{fig:DoS-both-strategies}
\end{figure}

In addition to studying the approximation ratio for the distance strategy, we compare in Fig.~\ref{fig:DoS-both-strategies} the evolution with $w$ of the approximation ratio distribution, for both strategies.
This data is for the \ac{CL} of the same random graph used in Sec.~\ref{sec:deselection}with the same parameters also, except for the values of $w$, with $1/w \in \{ 1, 3, 5, 14 \}$.
The black lines correspond to the average approximation ratio obtained when sampling at random a configuration in an energy window $[E_{\rm MWIS}, E_{\rm MWIS} + \Delta E]$.

Here, the value $w = 1$ is particularly interesting, since in the literature $w$ is typically restricted to $w < 1/2$~\cite{nguyen-2023, bombieri-2024-CL-gap}.
This constraint ensures that the weights defined in Eq.~\eqref{eq:weight-profile-chain} are all positive, and thus that any defect in the antiferromagnetic ordering has a positive energy cost.
With $w \ge 1/2$, some defective configurations become energetically favored compared to interpretable configurations, so that the \ac{CL} ground state (and the low energy states) is in general not interpretable.
Nevertheless, an interpretation strategy can still correct them and provide an approximation ratio $r=1$. 
This is what we observe in Fig.~\ref{fig:DoS-both-strategies}, where states with $\Delta E \lesssim 1/2-w $ are still interpreted correctly by both strategies.

For the deselection strategy, this can be explained as follows.
Suppose one starts from an interpretable \ac{CL} configuration and add defects at random.
If only domain walls are considered, adding a defect to a selected chain (in the sense of App.~\ref{app:algo_NIS}, that is when the chain is antiferromagnetically ordered and its endpoint with weight $1/2+w$ is selected) costs an energy $1/2 + w$, while adding a defect to an unselected chain costs an energy $1/2 - w$.
By construction, for $0 \le w \le 1/2$, both cost energy, so that interpretable states are favored.
But for $w > 1/2$, adding a defect to an unselected chain is energetically favorable, while adding a defect to a selected chain is energetically prohibitive.
In this $w$ regime, low energy states therefore contain defects in chains that would otherwise be unselected.
The strategy of deselecting defective chains then allows to recover good solutions to the original MIS problem despite the presence of potentially many defects.
Finally, we point out that this argument also holds for gadget-induced defects with a higher energy cost (like $3/2 - w$).

\subsection{Effect of the parameter $w$}

Our computations use random unweighted graphs $G$ generated with $N \in \{3, \dots, 7\}$ vertices and with an edge probability $p = 1/2$.
The ordering of the vertices, on which their \ac{CL} embedding graph depends, is thus chosen at random.
These graphs have in average 5 (for $N = 3$) to 22 ($N = 7$) \ac{IS}s, and a MIS size $\vert \text{MIS} \vert$ going from 1 to 4.
In each resulting \ac{CL} graph, the embedded \ac{DoS} thus spans a quite restricted energy window with $\vert \text{MIS} \vert+1$ peaks that are evenly spread across $\Delta E \in [0, 2w\vert \text{MIS} \vert]$.
An interesting consequence is that some visual features of the \ac{DoS} that depend on $w\cdot \vert \text{MIS} \vert$ can be deduced for large graphs (high $\vert \text{MIS} \vert$) from small graphs, simply by tuning $w$.

\begin{figure}
    \centering
    \includegraphics
    {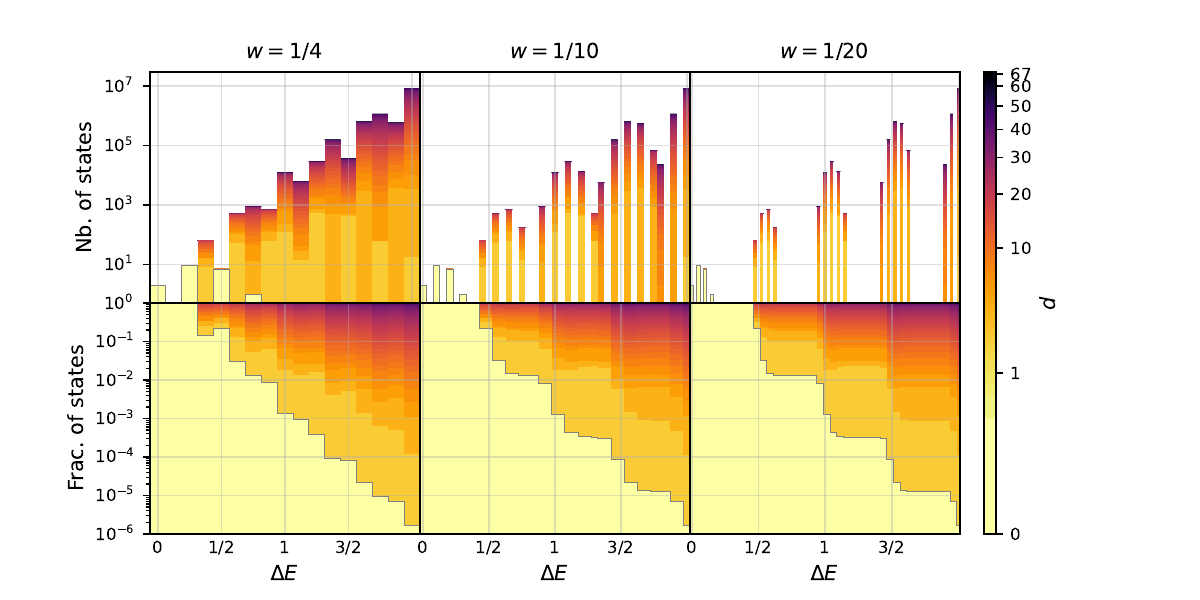}
    \caption{
    Effect of the weight bias $w$  for three values of $1/w \in [4, 10, 20]$ on a random \ac{CL} graph with initial size $N = 6$.
    \textit{(Top)} \ac{DoS} of the \ac{CL} graph with the usual weight profile $W_w$. 
    When $w \longrightarrow 0$, interpretable states corresponding to sub-optimal solutions in the initial graph are not energetically penalized.
    As a result, the energy of any configuration is given by the energy of its defects, which is approximately an integral multiple of $1/2$.
    \textit{(Bottom)}
    Corresponding distribution of the fraction of $d$-interpretable states after cumulating with respect to $E$, like in Fig.~\ref{fig:CL-and-paths}.
    }
    \label{fig:effect-w}
\end{figure}

The regime of particularly small weight bias is illustrated in Fig.~\ref{fig:effect-w} and is the basis of our analysis with respect to the domain wall number $l$ in the main text.
Recall that in the case of path graphs, under our weighting convention, a domain wall always induces a weight reduction by $1/2 \pm w$.
This remains true in the \ac{CL} graph, except for some gadget induced-defects.
Indeed, due to the presence of vertices with weight $2$ in the non-crossing gadget shown on Fig.~\ref{fig:CL-construction}~\textit{b)}, some defects are associated with a weight decrease $3/2 \pm w$.
Since $3/2$ is itself a multiple of $1/2$, any configuration has an energy which is an integral multiple of $1/2$ in the limit $w \longrightarrow 0^+$.

To sum up, at small values of $w$, the interpretable states are located in a narrow energy window above the \ac{MWIS}.
The width of this embedded \ac{DoS} scales linearly with $N$ for a broad class of random graphs~\cite{coja-2011}, but finding its exact width (or equivalently the \ac{MWIS} weight in the initial problem) is itself an NP-complete problem~\cite{karp-1972-reductibility-combinatorial-problems}.
The regime of small $w$ has the advantage of increasing the fraction $\tau_{d=0, \mathrm{CL}}(E)$ for small enough $E$, meaning that annealing outputs are more likely to be interpretable.
The downfall is that interpretable states with a poor approximation ratio are also more likely to be found, decreasing the quality of the output after interpretation.

On contrary, large values of $w$ penalize such sub-optimal interpretable solutions by endowing them with a high energy.
This different regime, however, is plagued by the small energy $1/2-w$ of defects, making non-interpretable configurations more likely to be found even at about $\simeq 1$ above the groundstate energy.

\subsection{Example of high-$d$ configuration}

\begin{figure}
    \centering
    \includegraphics[width=1.0\linewidth]{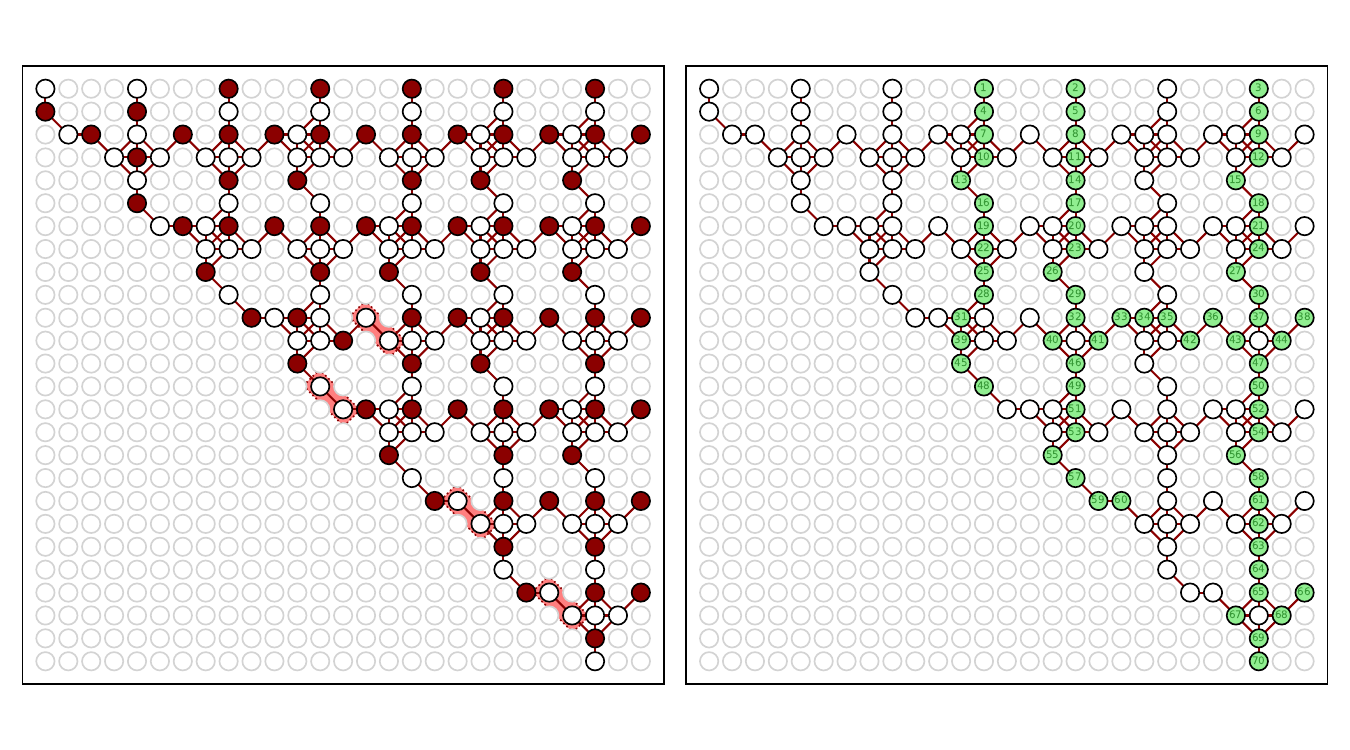}
    \caption{
    \textit{(Left)} 
    Non-interpretable configuration taken from the configurations shown in Fig.~\ref{fig:CL-and-paths}, here with the distance $d = 70$.
    Four domain walls are highlighted in light red, in good agreement with the energy $\Delta E \simeq 2$ of this configuration.
    \textit{(Right)}
    Example of 70 vertices that should be flipped to reach the closest interpretable configuration.
    }
    \label{fig:defective-config-CL}
\end{figure}

In Fig.~\ref{fig:CL-and-paths}, the highest distance $d$ that is achieved (in the probed energy range) is $d = 70$.
We illustrate in Fig.~\ref{fig:defective-config-CL} a configuration that achieves this distance \textit{(left pannel)} and highlight a possible choice of $d$ bitflips to interpret it \textit{(right panel)}.
Despite the hardness to interpret it, this configuration is located at a relatively low energy $E$ such that $\Delta E \simeq 2$.

\subsection{Example of harmful and harmless defects}

Here we introduce the simplest example of harmful and harmless defects, following the discussion in Sec.~\ref{sec:deselection}.
Let the initial graph by
\begin{equation}
    G = \Big(V = \{v_1, v_2\}, 
    \; E = \{ \{v_1, v_2\} \}\Big),
\end{equation}
namely the unweighted graph with two adjacent vertices.
Fig.~\ref{fig:harmless-harmfull-defects}~\textit{(Left)} represents one of the two \ac{MWIS}s of the corresponding \ac{CL} embedding.
We recall that the profile weight is a generalization of Eq.~\eqref{eq:weight-profile-chain}, so that both uppermost vertices contribute to an energy $-1/2-w$ if selected, while the rightmost and the lowermost vertices both contribute by an energy $-1/2+w$.

Without loss of generality, we consider this configuration represents the initial \ac{IS} $\{ v_1 \}$, and thus that $v_1$ is embedded by the leftmost chain.
If a single defect is added to this selected chain, as showed in Fig.~\ref{fig:harmless-harmfull-defects}~\textit{(Right)}, the rightmost vertex of the chain becomes selected.
Thus the resulting defective configuration has a higher energy $E = E_\mathrm{MWIS} + 1-2 -w$.
However, if we try to interpret it with the deselection strategy, which consists in ignoring defective chain, then it only remains the unselected chain, so the output approximation ratio is $r = 0$.
This justifies the denomination of harmful defects for defects that can be seen as added on selected chains.
Conversely, if the defect where to be added on the unselected chain, like in Fig.~\ref{fig:harmless-harmfull-defects}~\textit{(Middle)}, then the deselection strategy would produce the same $r = 1$ as if there were no defect at all.

\begin{figure}
    \centering
    \includegraphics[width=1\linewidth]{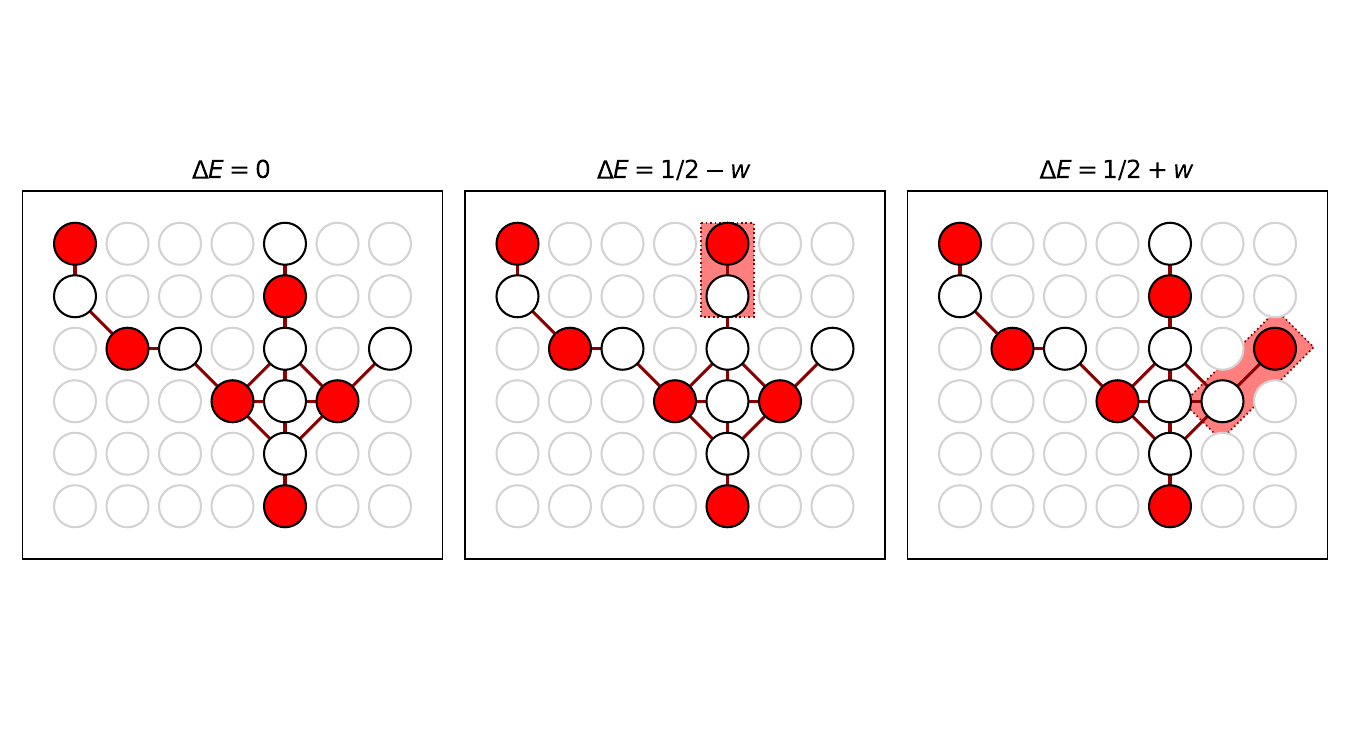}
    \caption{
    \textit{(Left)}
    \ac{MWIS} of the \ac{CL} embedding of a graph with two connected vertices.
    \textit{(Middle)}
    Configuration obtained by adding one harmless defect to the unselected chain of the MWIS, leaving the approximation ratio unchanged when using the deselection strategy.
    \textit{(Right)}
    Configuration with a harmful defect added to the selected chain of the MWIS, resulting in a loss of approximation ratio under the deselection strategy.
    }
    \label{fig:harmless-harmfull-defects}
\end{figure}

\subsection{List of possible defects}
\label{app:list-possible-defects}

\begin{figure}
    \centering
    \includegraphics[width=1\linewidth]{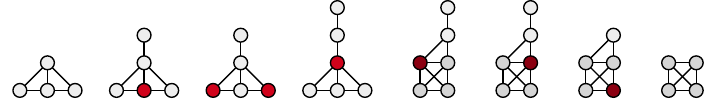}
    \caption{Eight gadget sub-configurations representing all possible gadget-induced defects, up to rotational symmetries (for the non-crossing gadget), diagonal symmetries and  deselection of some vertices.}
    \label{fig:gadget-induced-defects}
\end{figure}

In the main text, we systematically referred to defects in the \ac{CL} termed as domain wall-like defects.
Some defects are indeed domain walls, like the ones shown in Fig.~\label{fig:harmless-harmfull-defects} (\textit{middle and right}) and the leftmost defect (\textit{red highlight)} in Fig.~\ref{fig:defective-config-CL}.
However, the notion of domain walls is not precisely defined in the gadgets due to their more complex connectivity.
For the sake of completeness, we provide in Fig.~\ref{fig:gadget-induced-defects} the exhaustive list of gadget-induced defects.
The list is shortened as follows:
\begin{itemize}
    \item only defects that are maximal for the inclusion are represented; other defects can thus be obtained by deselecting one or more defects;
    \item when, in the non-crossing gadget, several defective configurations are obtained one from another through 90, 180 or 270 degrees rotations, only one is shown;
    \item when, in any of both gadgets, two defective configurations are identical up to a symmetry along the diagonal, only one is shown.
\end{itemize}

\section{Subdivision embedding} \label{app:subd-embedding}

\subsection{2D and 3D graph subdivisions}

We discuss in this section a different kind of embedding than the \ac{CL} scheme.
Given a graph, an \textit{even-edge subdivision} is the operation of inserting an even number of vertices $v_1, \dots, v_{2M}$ in a given edge.
The new edges are $(v_l, v_1), (v_1, v_2), \dots, (v_{2M}, v_r)$, so that the initial edge has been replaced by an even-length path subgraph.

Even-edge subdivisions have been used to embed non-planar or non-\ac{UD} graphs into \ac{UD} graphs \cite{pichler-2018-experimental, pichler-2018-computational, ebadi-2022}.
This techniques can be generalized in 3D where the analogous to \ac{UD} graphs is the class of \textit{unit-ball} graphs \cite{dalyac-2023-phd-thesis, kim-2022-3d-rydberg-wires} (for which efficient approximation schemes also exist \cite{bonamy-2018-eptas-UBMIS}).

Because even-length path subgraphs are used here to embed the state of a pair $(v_l, v_r)$ of adjacent vertices in $G$, the two antiferromagnetic configurations are not in one-to-one correspondence with the three possible configurations for $(v_l, v_r)$.
The generally used convention is thus that the antiferromagnetic states of the inserted path only encode the case when one vertex at the endpoints is selected.
When the path has one domain wall, the configuration can be considered as frustrated and discarded~\cite{kim-2022-3d-rydberg-wires}, or both endpoints can be considered to be unselected~\cite{pichler-2018-computational}.
In the following we adopt an intermediate stance where interpretation is done by contracting the subdivided edges to retrieve the embedded graph.
Independence violations are discarded, typically by deselecting the vertices in violation.
Note that in any case, the formalization of this embedding relies on a multi-valued embedding map \textit{f} with respect to the criteria discussed in the main text, whence a larger domain of definition for the interpretation map $f^{-1}$.

\subsection{Limitations of interpretability and weight profile}

\begin{figure}
    \centering
    \includegraphics[width=0.8\linewidth]{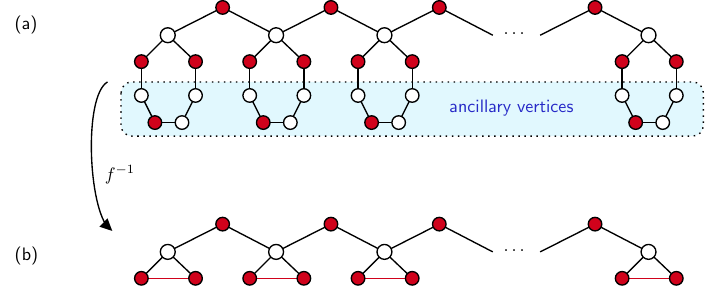}
    \caption{Initial graph \textit{(b)} and embedding graph \textit{(a)} in a pathological example of even-edge subdivision. If left unweighted, the embedding graph has a highly degenerate \ac{MIS}. Upon interpretation by edge contraction \textit{(b)}, the average number of independence violations \textit{(red edges)} scales linearly with the vertex count of the initial size.}
    \label{fig:pathological-example-odd-cycle}
\end{figure}

Here, we point out that subdivision embedding schemes are prone to independence violation upon interpretation in unweighted subdivided graphs.
An example of this issue is given in Fig.~\ref{fig:pathological-example-odd-cycle}.
The instance of initial graph \textit{(top)} has the particularity of having multiple odd-length cycle subgraphs.
(The fact that it is \ac{UD} and connected bears no importance here.)
By performing arbitrary even-edge subdivisions, one obtains an embedding graph where the number of odd-length cycle subgraphs remains identical \textit{(bottom)}.
However, odd-length cycles have a highly degenerate \ac{MIS} akin to the one in unweighted path graphs $P_{2N}$, with at most one domain wall.
Our example builds upon this property so that upon interpretation, subdivided edges can be interpreted into initial edges where both endpoints are selected, namely they lead to an independence violation that must be discarded.
Furthermore, if an \ac{MIS} of this embedding graph is taken at random, then each cycle has a non-zero probability (here 1/2) to have a domain wall and the average number of violations scales linearly with the graph size.
Note that the initial graph is connected, showing that odd-length cycles can give raise to independence violation even when embedded in more complex graph.
The initial graph is also \ac{UD} and with small maximum degree, for the clarity of representation.

This example generally proves that graphs with odd-length cycle subgraphs are not good candidates for unweighted even-edge subdivisions.
In theory, this issue can be avoided by adopting a weight profile like the $W_w$ from path graph that energetically penalizes domain walls.
This solution faces two obstacles in practice.
First, as shown in the main text, imperfect \ac{UD}-\ac{MWIS} optimization still raises interpretability issues.
In the case of subdivided graphs, this is  expected to yield states with multiple independence violations akin to the example of Fig.~\ref{fig:pathological-example-odd-cycle}.

A second obstacle is that the weight profile is itself constrained by experimental needs.
For Rydberg atoms, the van der Waals interaction has long-range effects that are not taken into account in the model of \ac{UD} graphs. 
In particular, higher detunings have to be applied to the atoms at the corner of the lattice or with a high number of neighbours~\cite{pichler-2018-computational, dalyac-2023-phd-thesis, Oliveira-2024}.
This modification aims at counteracting the long-range repulsion between the second-nearest neighbors.
However, according to the previous discussions, this causes an adversarial effect with respect to the weight profile $W_w$.
This weight profile penalizes defects by having lower weights on the endpoints of the antiferromagnetic chains than on the inner vertices.
This suggests that subdivision embeddings face the issue of degenerate domain walls, with constraints on the weight profile that would otherwise be used to alleviate these defects.

\section{Mitigation} \label{app:mitigation}

In this section, we study a weight profile on the \ac{CL} embedding that is intended to penalize the localization of defects in the bulk of the \ac{CL} graph.
The motivation of this approach is to extend the defect penalization from the weight profile $W_w$ introduce in the main text, which worked by lowering the weights on the endpoints of path graphs.
However, this modification has to satisfy an energy rescaling condition to ensure it describes the realistic constraint that the sum of all weights (or detunings in the case of Rydberg atoms) is a constant.
We will see the modified weight profile has unwanted effects because of this condition, but that it is beneficial to the interpretation if the energy budget is increased.

\subsection{Modified weight profiles}

\subsubsection{Case of paths graphs}

Let $P_{2N}$ be some path graph with vertices $v_1, \dots, v_{2N}$ and edges between $v_i, v_{i+1}$.
For each vertex $v_i$, we introduce its \textit{distance to the endpoints} $d_e(v_i) = \min(i, 2N-i) \in \mathbb{N}$.
We seek to modify the weight profile $W_w$ from Eq.~\eqref{eq:weight-profile-chain} in order to penalize the presence of domain walls with a high distance to the endpoints.
To do so, let us introduce the penalty function $\pi$ defined by
\begin{equation}
    \forall \mu \geq 0, \quad
    \forall n \in \mathbb{N}, \quad
    \pi(n ; \; \mu) = (n + 1)^\mu .
\end{equation}
A generalized weight profile $\tilde{W}_{w, \mu}$ to Eq.~\eqref{eq:weight-profile-chain} is then 
\begin{equation}
    \tilde{W}_{w, \mu}(v_i) 
    = 
    W_w(v_i) 
    \cdot 
    \pi\left( d_e(v_i) ; \; \mu \right). 
\end{equation}
This weight profile on $P_{2N}$ inherits from $d_e$ and $W_w$ the properties required for an embedding such as the weight ordering condition from Eq.~\eqref{eq:weight-ordering-preservation}.
In addition, the $\pi(n ; \; \mu)$ factor penalizes the localization of domain walls far from both endpoints of $P_{2N}$.

\subsubsection{Case of the \ac{CL} graph}

This scheme can be generalized to the \ac{CL} embedding at the cost of a more refined weight profile.
We still use the gadgets from Ref.~\cite{nguyen-2023} which are delimited as blocks of $4\times 4$ sites on a grid lattice.
The \ac{CL} graph can in this case be partitioned in $4\times 4$ blocks where some blocks do not correspond to gadgets but rather to pairs of vertices from the diagonal, horizontal and vertical boundaries.
For each block corresponding to a gadget or placed on the diagonal, we define its \textit{block distance to the endpoints} $b_e \in \mathbb{N}$ as proportional to the distance of this block to the two blocks located at the endpoints of the diagonal in the \ac{CL} graph.
A block distance $b_e = 0$ is assigned to all remaining blocks, which are either on the vertical or on the horizontal boundary of the \ac{CL}.
Each vertex $v_i$ being located in a block with a well-defined bock distance $b_e$, we use the notation $b_e(v_i)$ to denote this mapping.
The weight profile $W_w$, this time over the \ac{CL} graph, is then modulated to define a new profile $\tilde{W}_{w, \mu}$ using the block distance:
\begin{equation}
    \tilde{W}_{w, \mu}(v_i) 
    = 
    W_w(v_i) 
    \cdot 
    \pi\left( b_e(v_i) ; \; \mu \right). 
\end{equation}
The weight ordering property and the optimality of the embedded \ac{MWIS}s remain guaranteed.
Indeed, since all vertex in each gadget are modulated by the same factor, the 3 or 4 interpretable sub-configurations in each gadget have the same weight irrespective of $\mu$.
The fact that $\pi\left( b_e(v) ; \; \mu \right) = \pi\left(0 ; \; \mu \right) = 1$ for any vertex $v$ with $W_w(v) = 1/2 \pm w$ ensures that all the weight biases undergo the same modulation.
(Note that other weight profiles with different weights \textit{inside} the gadgets could also be engineered.)
Eventually, this weight profile can be seen as a generalization of the one from Ref.~\cite{nguyen-2023}, which is recovered by taking $\mu = 0$.

\subsection{Results} \label{app:mitigation-results}

We study the low-energy \ac{DoS} for $\mu \in [0, 0.2, 0.4]$ in one of our random \ac{CL} graphs with initial size $N = 5$.
The sum of the weights $\sum_{v_i \in G_\mathrm{CL}}\tilde{W}_{w, \mu}(v_i) $ in the graph increases with $\mu$.
Since this quantity would be proportional to the sum of all detunings used with an array of Rydberg atoms, some rescaling is needed to ensure we model different weight profiles with the same amount of energy supplied through the detunings.
We do so by studying the low-energy \ac{DoS} in a window $[E_{\text{MWIS}}, E_\mu]$ such that the ratio between $E_\mu-E_{\text{MWIS}}$ and $\sum_{v_i \in G_\mathrm{CL}}\tilde{W}_{w, \mu}(v_i)$ is the same for all values of $\mu$, with $E_{\mu=0} = 2$.

\begin{figure}
    \centering
    \includegraphics{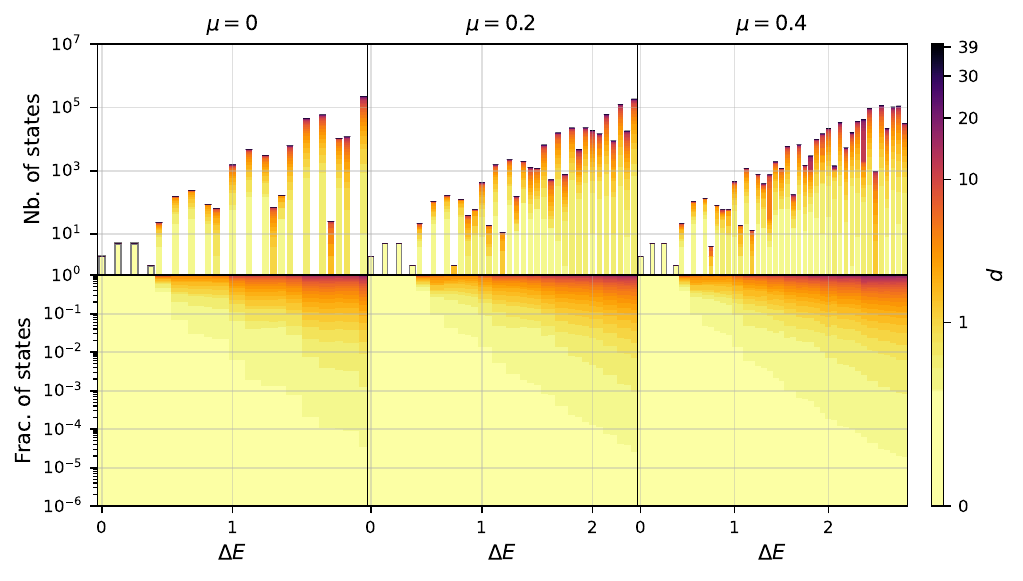}
    \caption{
        \textit{(Top)} 
        Low-energy \ac{DoS} for three set of parameters of the modified weight profile $\tilde{W}_{w, \mu}$ for a random \ac{CL} graph with initial size $N = 5$.
        The indicated energies correspond to the energies in the \ac{CL} before renormalization by the total energy (sum of all detunings), which depends on the weight profile.
        However the probed window is chosen so that the same fraction of the total energy is shown in all plots.
        \textit{(Bottom)}
        Corresponding distribution of the fraction of $d$-interpretable states after cumulating with respect to the energy.
    }
    \label{fig:DoS_modified_weight_profile_nu1}
\end{figure}

In Fig.~\ref{fig:DoS_modified_weight_profile_nu1}, we plot the three resulting low-energy \ac{DoS}.
We see that the modified weight profile slightly bends the curve of the fraction $\tau_{d, \mathrm{CL}}(E)$, at least for small distances $d$.
After the energy renormalization, both the value of $\tau_{d, \mathrm{CL}}(E)$ and the lowest-energy of non-interpretable states decrease when $\mu$ increases.
These observations suggest the energy rescaling indirectly favors defects with a low $b_e$ in a way that dwarfs the benefits from penalizing high-$b_e$ defects, ultimately making the modified weight profile slightly detrimental to the interpretation of approximate \ac{UD}-\ac{MWIS} solutions.

\begin{figure}
    \centering
    \includegraphics[width=1\linewidth]{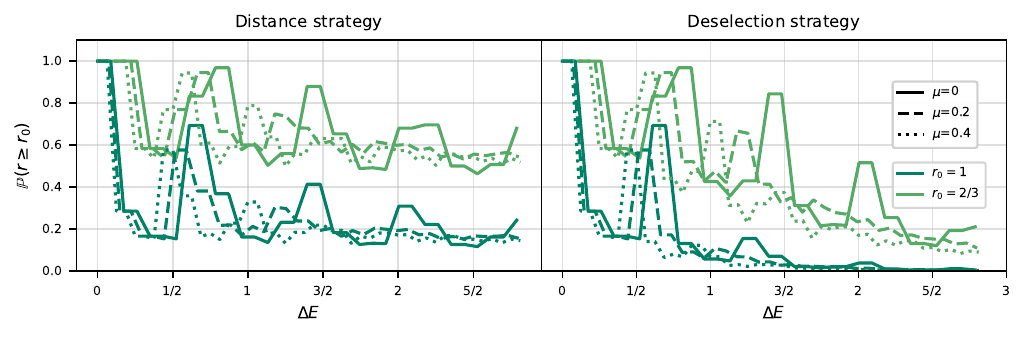}
    \caption{Probabilities of obtained high approximation ratios when uniformly sampling a random configuration of energy at most $\Delta E$, with the unscaled energy scale corresponding to the one for $\mu = 0.4$ in Fig.~\ref{fig:DoS_modified_weight_profile_nu1}.}
    \label{fig:comparison-strategies-nonzero-mu}
\end{figure}

In addition to the distribution of $d$, we plot in Fig.~\ref{fig:comparison-strategies-nonzero-mu} the probability $\mathbb{P}(r \ge r_0)$ to obtain at least some approximation ratio $r_0 \in \{ 2/3, 1\}$ when sampling at random a configuration below a given energy $\Delta E$.
Due to the global energy rescaling, the peaks of the different curves are not exactly aligned.
However we observe that increasing $\mu$ smooths the distribution, and that the distance strategy still yields higher probabilities than the deselection strategy (not shown on the plot) like in the main text.

\begin{figure}
    \centering
    \includegraphics{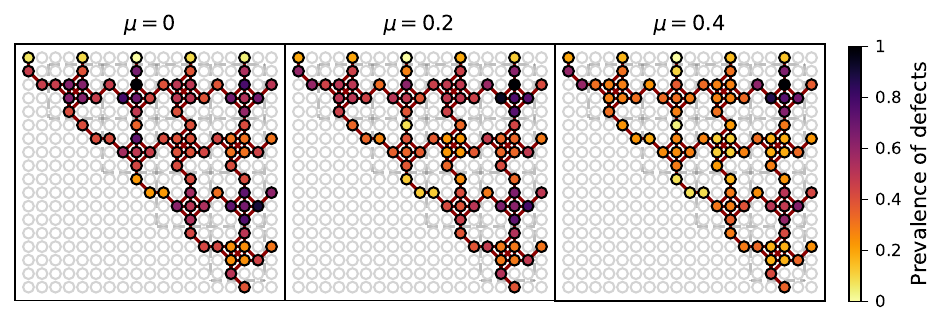}
    \caption{Distribution of the defects in a \ac{CL} graph corresponding to the probed configurations from Fig.~\ref{fig:DoS_modified_weight_profile_nu1}.
    We observe that increasing $\mu$ leads to relatively less defects located at the center of the \ac{CL}.
    }
    \label{fig:prevalence_defects_modified_weight_profile}
\end{figure}

\begin{figure}
    \centering
    \includegraphics[width=1.0\linewidth]{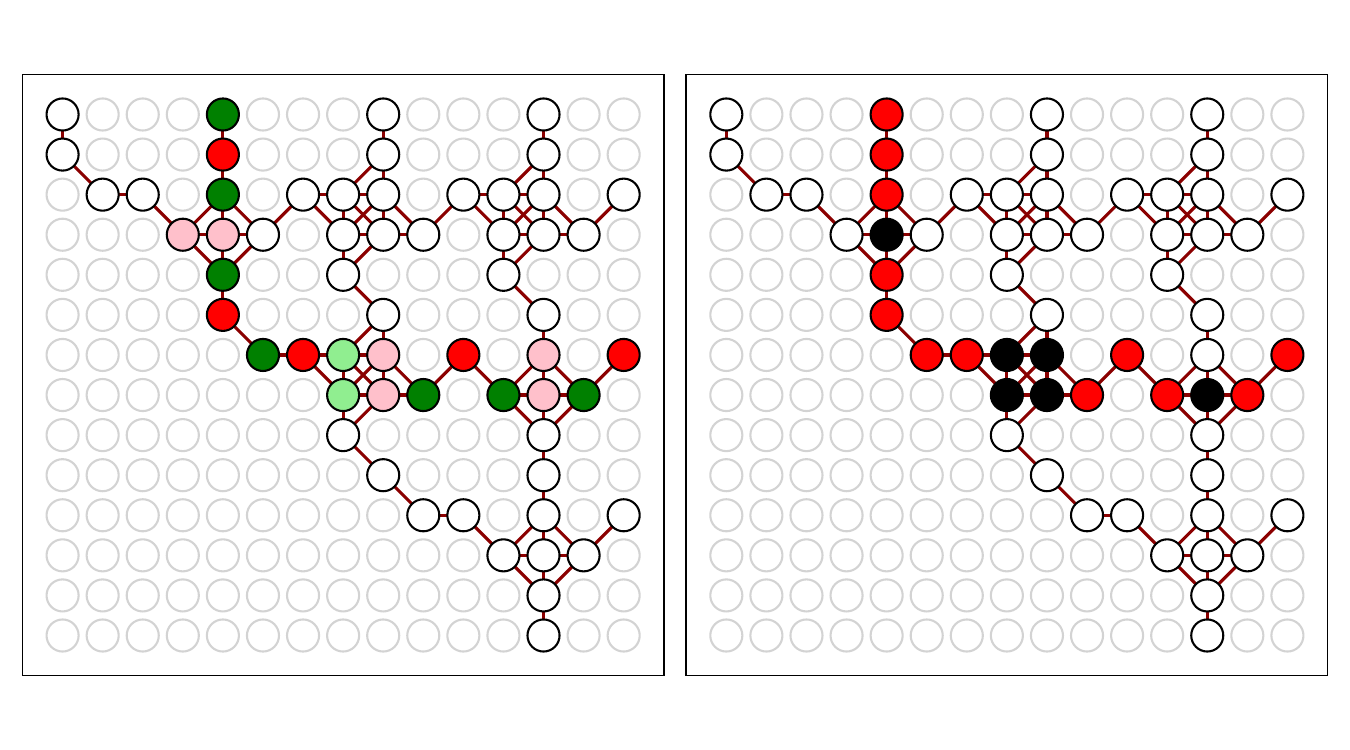}
    \caption{
    \textit{(Left)} 
    Example of a chain as defined in App.~\ref{app:mitigation-results}, where gadgets are inserted by splitting some vertices (lighter colors). Thanks to this, it is possible to define a parity along the chain (red and green colors), and to identify any defect as the succession of two different colors with no selected vertex.
    \textit{(Right)}
    Example of path nodes (red) and gadget nodes (black) as defined in App~\ref{app:algo_NIS}. The \ac{CL} can be partitioned between the path nodes of all $N$ chains and the gadgets nodes. 
    In an interpretable configuration, the path nodes have an antiferromagnetic ordering specified by wether the corresponding path is selected or not.
    }
    \label{fig:def-paths-chains}
\end{figure}

Despite the unwanted variation in the distribution of $\tau_{d, \mathrm{CL}}(E)$, we verify that the weight profile $\tilde{W}_{w, \mu}$ does penalize defects with a high $b_e$ as follows.
Given a configuration $S$ in the \ac{CL} graph and a vertex $v_i$, we denote by $\xi(v_i \vert S) \in \mathbb{N}$ the number of domain walls located at $v_i$ (and at least one of its neighbors).
This is done by looking for two successive unselected vertices along each chain.
In the crossing (resp. non-crossing) gadgets, this approach readily generalizes by considering that 1 (resp. 2) vertex along the chain undergoes a vertex splitting, so that the corresponding defect is identified as 3 (resp. 4) unselected vertices.
An example is given in Fig.~\ref{fig:def-paths-chains}.
(Note that if two or more defects are adjacent one vertex $v_i$ can have $\xi(v_i \vert S) > 1$.)
The value of $\xi(v_i \vert S)$ is then summed across all probed configurations, all other parameters being the same as in Fig.~\ref{fig:DoS_modified_weight_profile_nu1}.
For each $\mu$, the result is linearly remapped so that its extremal values across all $v_i$ are 0 and 1.
The remapped value is plotted in Fig.~\ref{fig:prevalence_defects_modified_weight_profile} and can be understood as a measure of the prevalence of defects across the \ac{CL} graph.
As expected, we observe that for higher values of $\mu$ than the default $\mu = 0$, vertices with a high block distance $b_e$ are energetically penalized.

\section{Finding the nearest interpretable state as a QUBO problem}
\label{app:algo_NIS}

Here we describe the task of finding the nearest interpretable state of a \ac{CL} configuration $S$ as a \ac{QUBO} problem closely related to an MWIS problem on a subgraph of the original graph.

We want to minimize the distance $d$ between a fixed configuration $S$ and any interpretable configuration $S_{\rm int}$.
This distance can be decomposed into contributions from path nodes (paths are indexed by $p$) and gadget nodes (gadgets are indexed by $(p, q)$)
\begin{equation}
    d = \sum_p d_p + \sum_{p > q} d_{pq}.
\end{equation}
In this appendix, we use the following definitions.
Gadget nodes are either i) the central node with 4 edges in the crossing-with-edge gadget, or ii) the central 4-clique in the crossing-without-edge gadget.
Path nodes are all the other nodes and they can be grouped by path in a straightforward way.
An example is shown in Fig.~\ref{fig:def-paths-chains} \textit{(right)}, where all vertices along a path are colored depending on wether they are considered as path nodes \textit{(red)} or gadget nodes \textit{(black)}.
In an interpretable state, path nodes selection is determined only by the status (selected/unselected) of the path in which they belong, any ambiguity has been moved out to gadget nodes.
We denote $n_p = 0$ if path $p$ is unselected in $S_{int}$ and $n_p=1$ if it is selected.

For each path $p$, denote $d_p^0$ ($d_p^1$) the distance generated from $S$ by deselecting (selecting) path nodes along path $p$ (excluding gadget nodes).
Then we have
\begin{equation}
    d_p = d_p^0 + (d_p^1 - d_p^0) n_p.
\end{equation}

In an interpretable configuration, each gadget can be in at most 4 states, corresponding to all possibilities of selecting or deselecting adjacent paths, which we label by $00, 01, 10$ and $11$.
Therefore, for a given $S$, each gadget $(p, q)$ can contribute to one of only 4 possible values of distance, which we denote $d_{pq}^{ab}$ for $a, b \in \{0, 1\}$.
In fact, crossing-with-edge gadgets can only be in two states, since they are a single node in our definition.
For these, we will still use the same labels and keep in mind that $d_{pq}^{01} = d_{pq}^{10}$ and set $d_{pq}^{11} = 0$ since the case where both $p$ and $q$ are selected does not exist in an interpretable state.
The contribution of an arbitrary gadget $(p, q)$ to the distance is therefore
\begin{equation}
    d_{pq}
    = 
    d_{pq}^{00} [1 - n_p] [1 - n_q]
    + d_{pq}^{01} [1 - n_p] n_q
    + d_{pq}^{10} n_p [1 - n_q]
    + d_{pq}^{11} n_p n_q.
\end{equation}
For crossing-with-edge gadgets this simplifies to
\begin{align}
    d_{pq} = d_{pq}^{00}  + [n_p + n_q] [d_{pq}^{01} - d_{pq}^{00}],
\end{align}
using the fact that with such gadget, $n_p n_q$ is always zero.

All together, we see that $d$ is a quadratic form
\begin{equation}
    d = C + \sum_p D_p n_p + \sum_{p > q} Q_{pq} n_p n_q
\end{equation}
which must be minimized under the constraint that $\{n_p\}$ should form an independent set of the original graph $G$.
This constraint can be formally introduced with a large extra term $\bar Q_{pq}$ for each edge $(p,q)$ of $G$, namely 
\begin{equation}
    d = C + \sum_p D_p n_p + \sum_{(p,q) \in \bar E} Q_{pq} n_p n_q + \sum_{(p,q) \in E} \bar Q_{pq} n_p n_q
\end{equation}
where $E$ is the set of edges of $G$, $\bar E$ its complementary, and $\bar Q_{pq} \gg Q_{p'q'}, D_{p'}$.
In this form this is a \ac{QUBO} problem.

We now show the proximity of this problem to an instance of \ac{MWIS}.
Observing that the quadratic term $Q$ only comes from crossing-without-edge gadgets, it can be seen as a perturbation compared to the stronger independence constraint.
If we ignore it, the task become equivalent to maximizing
\begin{equation}
    f = \sum_p (d_p^0 - d_p^1) n_p
\end{equation}
under the independence constraint.
Paths $p$ such that $d_p(0) \le d_p(1)$ can be ignored, since they must be unselected in the optimal state (if it was not the case, then deselecting it would increase $f$ without breaking the independence constraint).
We can thus remove these paths from the optimization and we are left with positive weights only.
Therefore, this problem is equivalent to finding an \ac{MWIS} on the subgraph of the initial graph composed of those nodes where $d_p(0) > d_p(1)$, with weights $d_p(0) - d_p(1)$.
Informally, $d_p(0) > d_p(1)$ means that the path $p$ is closer to being selected than unselected in $S$.
Interestingly, this tells us that, in the search for the distance, paths that are closer to being unselected can be set to unselected right away, if one is happy to ignore the contribution from crossing-without-edge gadgets.


\begin{thebibliography}{44}%
\makeatletter
\providecommand \@ifxundefined [1]{%
 \@ifx{#1\undefined}
}%
\providecommand \@ifnum [1]{%
 \ifnum #1\expandafter \@firstoftwo
 \else \expandafter \@secondoftwo
 \fi
}%
\providecommand \@ifx [1]{%
 \ifx #1\expandafter \@firstoftwo
 \else \expandafter \@secondoftwo
 \fi
}%
\providecommand \natexlab [1]{#1}%
\providecommand \enquote  [1]{``#1''}%
\providecommand \bibnamefont  [1]{#1}%
\providecommand \bibfnamefont [1]{#1}%
\providecommand \citenamefont [1]{#1}%
\providecommand \href@noop [0]{\@secondoftwo}%
\providecommand \href [0]{\begingroup \@sanitize@url \@href}%
\providecommand \@href[1]{\@@startlink{#1}\@@href}%
\providecommand \@@href[1]{\endgroup#1\@@endlink}%
\providecommand \@sanitize@url [0]{\catcode `\\12\catcode `\$12\catcode `\&12\catcode `\#12\catcode `\^12\catcode `\_12\catcode `\%12\relax}%
\providecommand \@@startlink[1]{}%
\providecommand \@@endlink[0]{}%
\providecommand \url  [0]{\begingroup\@sanitize@url \@url }%
\providecommand \@url [1]{\endgroup\@href {#1}{\urlprefix }}%
\providecommand \urlprefix  [0]{URL }%
\providecommand \Eprint [0]{\href }%
\providecommand \doibase [0]{https://doi.org/}%
\providecommand \selectlanguage [0]{\@gobble}%
\providecommand \bibinfo  [0]{\@secondoftwo}%
\providecommand \bibfield  [0]{\@secondoftwo}%
\providecommand \translation [1]{[#1]}%
\providecommand \BibitemOpen [0]{}%
\providecommand \bibitemStop [0]{}%
\providecommand \bibitemNoStop [0]{.\EOS\space}%
\providecommand \EOS [0]{\spacefactor3000\relax}%
\providecommand \BibitemShut  [1]{\csname bibitem#1\endcsname}%
\let\auto@bib@innerbib\@empty
\bibitem [{\citenamefont {Pichler}\ \emph {et~al.}(2018{\natexlab{a}})\citenamefont {Pichler}, \citenamefont {Wang}, \citenamefont {Zhou}, \citenamefont {Choi},\ and\ \citenamefont {Lukin}}]{pichler-2018-computational}%
  \BibitemOpen
  \bibfield  {author} {\bibinfo {author} {\bibfnamefont {H.}~\bibnamefont {Pichler}}, \bibinfo {author} {\bibfnamefont {S.-T.}\ \bibnamefont {Wang}}, \bibinfo {author} {\bibfnamefont {L.}~\bibnamefont {Zhou}}, \bibinfo {author} {\bibfnamefont {S.}~\bibnamefont {Choi}},\ and\ \bibinfo {author} {\bibfnamefont {M.~D.}\ \bibnamefont {Lukin}},\ }\href@noop {} {\bibfield  {journal} {\bibinfo  {journal} {arXiv preprint arXiv:1809.04954}\ } (\bibinfo {year} {2018}{\natexlab{a}})}\BibitemShut {NoStop}%
\bibitem [{\citenamefont {Pichler}\ \emph {et~al.}(2018{\natexlab{b}})\citenamefont {Pichler}, \citenamefont {Wang}, \citenamefont {Zhou}, \citenamefont {Choi},\ and\ \citenamefont {Lukin}}]{pichler-2018-experimental}%
  \BibitemOpen
  \bibfield  {author} {\bibinfo {author} {\bibfnamefont {H.}~\bibnamefont {Pichler}}, \bibinfo {author} {\bibfnamefont {S.-T.}\ \bibnamefont {Wang}}, \bibinfo {author} {\bibfnamefont {L.}~\bibnamefont {Zhou}}, \bibinfo {author} {\bibfnamefont {S.}~\bibnamefont {Choi}},\ and\ \bibinfo {author} {\bibfnamefont {M.~D.}\ \bibnamefont {Lukin}},\ }\href@noop {} {\bibfield  {journal} {\bibinfo  {journal} {arXiv preprint arXiv:1808.10816}\ } (\bibinfo {year} {2018}{\natexlab{b}})}\BibitemShut {NoStop}%
\bibitem [{\citenamefont {Ebadi}\ \emph {et~al.}(2022)\citenamefont {Ebadi}, \citenamefont {Keesling}, \citenamefont {Cain}, \citenamefont {Wang}, \citenamefont {Levine}, \citenamefont {Bluvstein}, \citenamefont {Semeghini}, \citenamefont {Omran}, \citenamefont {Liu}, \citenamefont {Samajdar}, \citenamefont {Luo}, \citenamefont {Nash}, \citenamefont {Gao}, \citenamefont {Barak}, \citenamefont {Farhi}, \citenamefont {Sachdev}, \citenamefont {Gemelke}, \citenamefont {Zhou}, \citenamefont {Choi}, \citenamefont {Pichler}, \citenamefont {Wang}, \citenamefont {Greiner}, \citenamefont {Vuletić},\ and\ \citenamefont {Lukin}}]{ebadi-2022}%
  \BibitemOpen
  \bibfield  {author} {\bibinfo {author} {\bibfnamefont {S.}~\bibnamefont {Ebadi}}, \bibinfo {author} {\bibfnamefont {A.}~\bibnamefont {Keesling}}, \bibinfo {author} {\bibfnamefont {M.}~\bibnamefont {Cain}}, \bibinfo {author} {\bibfnamefont {T.~T.}\ \bibnamefont {Wang}}, \bibinfo {author} {\bibfnamefont {H.}~\bibnamefont {Levine}}, \bibinfo {author} {\bibfnamefont {D.}~\bibnamefont {Bluvstein}}, \bibinfo {author} {\bibfnamefont {G.}~\bibnamefont {Semeghini}}, \bibinfo {author} {\bibfnamefont {A.}~\bibnamefont {Omran}}, \bibinfo {author} {\bibfnamefont {J.-G.}\ \bibnamefont {Liu}}, \bibinfo {author} {\bibfnamefont {R.}~\bibnamefont {Samajdar}}, \bibinfo {author} {\bibfnamefont {X.-Z.}\ \bibnamefont {Luo}}, \bibinfo {author} {\bibfnamefont {B.}~\bibnamefont {Nash}}, \bibinfo {author} {\bibfnamefont {X.}~\bibnamefont {Gao}}, \bibinfo {author} {\bibfnamefont {B.}~\bibnamefont {Barak}}, \bibinfo {author} {\bibfnamefont {E.}~\bibnamefont {Farhi}}, \bibinfo {author} {\bibfnamefont {S.}~\bibnamefont {Sachdev}},
  \bibinfo {author} {\bibfnamefont {N.}~\bibnamefont {Gemelke}}, \bibinfo {author} {\bibfnamefont {L.}~\bibnamefont {Zhou}}, \bibinfo {author} {\bibfnamefont {S.}~\bibnamefont {Choi}}, \bibinfo {author} {\bibfnamefont {H.}~\bibnamefont {Pichler}}, \bibinfo {author} {\bibfnamefont {S.-T.}\ \bibnamefont {Wang}}, \bibinfo {author} {\bibfnamefont {M.}~\bibnamefont {Greiner}}, \bibinfo {author} {\bibfnamefont {V.}~\bibnamefont {Vuletić}},\ and\ \bibinfo {author} {\bibfnamefont {M.~D.}\ \bibnamefont {Lukin}},\ }\href {https://doi.org/10.1126/science.abo6587} {\bibfield  {journal} {\bibinfo  {journal} {Science}\ }\textbf {\bibinfo {volume} {376}},\ \bibinfo {pages} {1209} (\bibinfo {year} {2022})},\ \Eprint {https://arxiv.org/abs/https://www.science.org/doi/pdf/10.1126/science.abo6587} {https://www.science.org/doi/pdf/10.1126/science.abo6587} \BibitemShut {NoStop}%
\bibitem [{\citenamefont {Kim}\ \emph {et~al.}(2022)\citenamefont {Kim}, \citenamefont {Kim}, \citenamefont {Hwang}, \citenamefont {Moon},\ and\ \citenamefont {Ahn}}]{kim-2022-3d-rydberg-wires}%
  \BibitemOpen
  \bibfield  {author} {\bibinfo {author} {\bibfnamefont {M.}~\bibnamefont {Kim}}, \bibinfo {author} {\bibfnamefont {K.}~\bibnamefont {Kim}}, \bibinfo {author} {\bibfnamefont {J.}~\bibnamefont {Hwang}}, \bibinfo {author} {\bibfnamefont {E.-G.}\ \bibnamefont {Moon}},\ and\ \bibinfo {author} {\bibfnamefont {J.}~\bibnamefont {Ahn}},\ }\href@noop {} {\bibfield  {journal} {\bibinfo  {journal} {Nature Physics}\ }\textbf {\bibinfo {volume} {18}},\ \bibinfo {pages} {755} (\bibinfo {year} {2022})}\BibitemShut {NoStop}%
\bibitem [{\citenamefont {de~Oliveira}\ \emph {et~al.}(2024)\citenamefont {de~Oliveira}, \citenamefont {Diamond-Hitchcock}, \citenamefont {Walker}, \citenamefont {Wells-Pestell}, \citenamefont {Pelegrí}, \citenamefont {Picken}, \citenamefont {Malcolm}, \citenamefont {Daley}, \citenamefont {Bass},\ and\ \citenamefont {Pritchard}}]{Oliveira-2024}%
  \BibitemOpen
  \bibfield  {author} {\bibinfo {author} {\bibfnamefont {A.~G.}\ \bibnamefont {de~Oliveira}}, \bibinfo {author} {\bibfnamefont {E.}~\bibnamefont {Diamond-Hitchcock}}, \bibinfo {author} {\bibfnamefont {D.~M.}\ \bibnamefont {Walker}}, \bibinfo {author} {\bibfnamefont {M.~T.}\ \bibnamefont {Wells-Pestell}}, \bibinfo {author} {\bibfnamefont {G.}~\bibnamefont {Pelegrí}}, \bibinfo {author} {\bibfnamefont {C.~J.}\ \bibnamefont {Picken}}, \bibinfo {author} {\bibfnamefont {G.~P.~A.}\ \bibnamefont {Malcolm}}, \bibinfo {author} {\bibfnamefont {A.~J.}\ \bibnamefont {Daley}}, \bibinfo {author} {\bibfnamefont {J.}~\bibnamefont {Bass}},\ and\ \bibinfo {author} {\bibfnamefont {J.~D.}\ \bibnamefont {Pritchard}},\ }\href {https://arxiv.org/abs/2404.02658} {\bibinfo {title} {Demonstration of weighted graph optimization on a rydberg atom array using local light-shifts}} (\bibinfo {year} {2024}),\ \Eprint {https://arxiv.org/abs/2404.02658} {arXiv:2404.02658 [quant-ph]} \BibitemShut {NoStop}%
\bibitem [{\citenamefont {Cazals}\ \emph {et~al.}(2025)\citenamefont {Cazals}, \citenamefont {Fran{\c{c}}ois}, \citenamefont {Henriet}, \citenamefont {Leclerc}, \citenamefont {Marin}, \citenamefont {Naghmouchi}, \citenamefont {Coelho}, \citenamefont {Sikora}, \citenamefont {Vitale}, \citenamefont {Watrigant} \emph {et~al.}}]{cazals-2025-identifying-hard-native-instances}%
  \BibitemOpen
  \bibfield  {author} {\bibinfo {author} {\bibfnamefont {P.}~\bibnamefont {Cazals}}, \bibinfo {author} {\bibfnamefont {A.}~\bibnamefont {Fran{\c{c}}ois}}, \bibinfo {author} {\bibfnamefont {L.}~\bibnamefont {Henriet}}, \bibinfo {author} {\bibfnamefont {L.}~\bibnamefont {Leclerc}}, \bibinfo {author} {\bibfnamefont {M.}~\bibnamefont {Marin}}, \bibinfo {author} {\bibfnamefont {Y.}~\bibnamefont {Naghmouchi}}, \bibinfo {author} {\bibfnamefont {W.~d.~S.}\ \bibnamefont {Coelho}}, \bibinfo {author} {\bibfnamefont {F.}~\bibnamefont {Sikora}}, \bibinfo {author} {\bibfnamefont {V.}~\bibnamefont {Vitale}}, \bibinfo {author} {\bibfnamefont {R.}~\bibnamefont {Watrigant}}, \emph {et~al.},\ }\href@noop {} {\bibfield  {journal} {\bibinfo  {journal} {arXiv preprint arXiv:2502.04291}\ } (\bibinfo {year} {2025})}\BibitemShut {NoStop}%
\bibitem [{\citenamefont {Karp}(1972)}]{karp-1972-reductibility-combinatorial-problems}%
  \BibitemOpen
  \bibfield  {author} {\bibinfo {author} {\bibfnamefont {R.~M.}\ \bibnamefont {Karp}},\ }\href@noop {} {\bibfield  {journal} {\bibinfo  {journal} {Complexity of computer computations}\ }\textbf {\bibinfo {volume} {1}},\ \bibinfo {pages} {85} (\bibinfo {year} {1972})}\BibitemShut {NoStop}%
\bibitem [{\citenamefont {Garey}\ and\ \citenamefont {Johnson}(1978)}]{garey-1978-MIS-strongly-NP-hard}%
  \BibitemOpen
  \bibfield  {author} {\bibinfo {author} {\bibfnamefont {M.~R.}\ \bibnamefont {Garey}}\ and\ \bibinfo {author} {\bibfnamefont {D.~S.}\ \bibnamefont {Johnson}},\ }\href {https://doi.org/10.1145/322077.322090} {\bibfield  {journal} {\bibinfo  {journal} {J. ACM}\ }\textbf {\bibinfo {volume} {25}},\ \bibinfo {pages} {499–508} (\bibinfo {year} {1978})}\BibitemShut {NoStop}%
\bibitem [{\citenamefont {Erlebach}\ \emph {et~al.}(2001)\citenamefont {Erlebach}, \citenamefont {Jansen},\ and\ \citenamefont {Seidel}}]{Erlebach-PTAS-UD-MWIS}%
  \BibitemOpen
  \bibfield  {author} {\bibinfo {author} {\bibfnamefont {T.}~\bibnamefont {Erlebach}}, \bibinfo {author} {\bibfnamefont {K.}~\bibnamefont {Jansen}},\ and\ \bibinfo {author} {\bibfnamefont {E.}~\bibnamefont {Seidel}},\ }in\ \href@noop {} {\emph {\bibinfo {booktitle} {Proceedings of the Twelfth Annual ACM-SIAM Symposium on Discrete Algorithms}}},\ \bibinfo {series and number} {SODA '01}\ (\bibinfo  {publisher} {Society for Industrial and Applied Mathematics},\ \bibinfo {address} {USA},\ \bibinfo {year} {2001})\ p.\ \bibinfo {pages} {671–679}\BibitemShut {NoStop}%
\bibitem [{\citenamefont {Matsui}(1998)}]{matsui-1998-approximating-MIS-on-UDG}%
  \BibitemOpen
  \bibfield  {author} {\bibinfo {author} {\bibfnamefont {T.}~\bibnamefont {Matsui}},\ }in\ \href@noop {} {\emph {\bibinfo {booktitle} {Japanese Conference on Discrete and Computational Geometry}}}\ (\bibinfo {organization} {Springer},\ \bibinfo {year} {1998})\ pp.\ \bibinfo {pages} {194--200}\BibitemShut {NoStop}%
\bibitem [{\citenamefont {Das}\ \emph {et~al.}(2015)\citenamefont {Das}, \citenamefont {De}, \citenamefont {Kolay}, \citenamefont {Nandy},\ and\ \citenamefont {Sur-Kolay}}]{Das-2015-approximating-MIS-on-UDG}%
  \BibitemOpen
  \bibfield  {author} {\bibinfo {author} {\bibfnamefont {G.~K.}\ \bibnamefont {Das}}, \bibinfo {author} {\bibfnamefont {M.}~\bibnamefont {De}}, \bibinfo {author} {\bibfnamefont {S.}~\bibnamefont {Kolay}}, \bibinfo {author} {\bibfnamefont {S.~C.}\ \bibnamefont {Nandy}},\ and\ \bibinfo {author} {\bibfnamefont {S.}~\bibnamefont {Sur-Kolay}},\ }\href {https://doi.org/https://doi.org/10.1016/j.ipl.2014.11.002} {\bibfield  {journal} {\bibinfo  {journal} {Information Processing Letters}\ }\textbf {\bibinfo {volume} {115}},\ \bibinfo {pages} {439} (\bibinfo {year} {2015})}\BibitemShut {NoStop}%
\bibitem [{\citenamefont {Nandy}\ \emph {et~al.}(2017)\citenamefont {Nandy}, \citenamefont {Pandit},\ and\ \citenamefont {Roy}}]{Nandy-2017-approximating-MIS-on-UDG}%
  \BibitemOpen
  \bibfield  {author} {\bibinfo {author} {\bibfnamefont {S.~C.}\ \bibnamefont {Nandy}}, \bibinfo {author} {\bibfnamefont {S.}~\bibnamefont {Pandit}},\ and\ \bibinfo {author} {\bibfnamefont {S.}~\bibnamefont {Roy}},\ }\href {https://doi.org/https://doi.org/10.1016/j.ipl.2017.07.007} {\bibfield  {journal} {\bibinfo  {journal} {Information Processing Letters}\ }\textbf {\bibinfo {volume} {127}},\ \bibinfo {pages} {58} (\bibinfo {year} {2017})}\BibitemShut {NoStop}%
\bibitem [{\citenamefont {Bonamy}\ \emph {et~al.}(2018)\citenamefont {Bonamy}, \citenamefont {Bonnet}, \citenamefont {Bousquet}, \citenamefont {Charbit},\ and\ \citenamefont {Thomass{\'e}}}]{bonamy-2018-eptas-UBMIS}%
  \BibitemOpen
  \bibfield  {author} {\bibinfo {author} {\bibfnamefont {M.}~\bibnamefont {Bonamy}}, \bibinfo {author} {\bibfnamefont {E.}~\bibnamefont {Bonnet}}, \bibinfo {author} {\bibfnamefont {N.}~\bibnamefont {Bousquet}}, \bibinfo {author} {\bibfnamefont {P.}~\bibnamefont {Charbit}},\ and\ \bibinfo {author} {\bibfnamefont {S.}~\bibnamefont {Thomass{\'e}}},\ }in\ \href@noop {} {\emph {\bibinfo {booktitle} {2018 IEEE 59th Annual Symposium on Foundations of Computer Science (FOCS)}}}\ (\bibinfo {organization} {IEEE},\ \bibinfo {year} {2018})\ pp.\ \bibinfo {pages} {568--579}\BibitemShut {NoStop}%
\bibitem [{\citenamefont {Serret}\ \emph {et~al.}(2020)\citenamefont {Serret}, \citenamefont {Marchand},\ and\ \citenamefont {Ayral}}]{serret-2020-benchmark}%
  \BibitemOpen
  \bibfield  {author} {\bibinfo {author} {\bibfnamefont {M.~F.}\ \bibnamefont {Serret}}, \bibinfo {author} {\bibfnamefont {B.}~\bibnamefont {Marchand}},\ and\ \bibinfo {author} {\bibfnamefont {T.}~\bibnamefont {Ayral}},\ }\href@noop {} {\bibfield  {journal} {\bibinfo  {journal} {Physical Review A}\ }\textbf {\bibinfo {volume} {102}},\ \bibinfo {pages} {052617} (\bibinfo {year} {2020})}\BibitemShut {NoStop}%
\bibitem [{\citenamefont {H{\aa}stad}(1999)}]{Hastad-1999-hardness-approximating-MIS}%
  \BibitemOpen
  \bibfield  {author} {\bibinfo {author} {\bibfnamefont {J.}~\bibnamefont {H{\aa}stad}},\ }\href {https://doi.org/10.1007/BF02392825} {\bibfield  {journal} {\bibinfo  {journal} {Acta Mathematica}\ }\textbf {\bibinfo {volume} {182}},\ \bibinfo {pages} {105 } (\bibinfo {year} {1999})}\BibitemShut {NoStop}%
\bibitem [{\citenamefont {Zuckerman}(2006)}]{zuckerman-2006-hardness-approximating-MIS}%
  \BibitemOpen
  \bibfield  {author} {\bibinfo {author} {\bibfnamefont {D.}~\bibnamefont {Zuckerman}},\ }in\ \href {https://doi.org/10.1145/1132516.1132612} {\emph {\bibinfo {booktitle} {Proceedings of the Thirty-Eighth Annual ACM Symposium on Theory of Computing}}}\ (\bibinfo  {publisher} {Association for Computing Machinery},\ \bibinfo {address} {New York, NY, USA},\ \bibinfo {year} {2006})\ p.\ \bibinfo {pages} {681–690}\BibitemShut {NoStop}%
\bibitem [{\citenamefont {Dalyac}(2023)}]{dalyac-2023-phd-thesis}%
  \BibitemOpen
  \bibfield  {author} {\bibinfo {author} {\bibfnamefont {C.}~\bibnamefont {Dalyac}},\ }\emph {\bibinfo {title} {Quantum many-body dynamics for combinatorial optimisation and machine learning}},\ \href@noop {} {Ph.D. thesis},\ \bibinfo  {school} {Sorbonne Universit{\'e}} (\bibinfo {year} {2023})\BibitemShut {NoStop}%
\bibitem [{\citenamefont {Nguyen}\ \emph {et~al.}(2023)\citenamefont {Nguyen}, \citenamefont {Liu}, \citenamefont {Wurtz}, \citenamefont {Lukin}, \citenamefont {Wang},\ and\ \citenamefont {Pichler}}]{nguyen-2023}%
  \BibitemOpen
  \bibfield  {author} {\bibinfo {author} {\bibfnamefont {M.-T.}\ \bibnamefont {Nguyen}}, \bibinfo {author} {\bibfnamefont {J.-G.}\ \bibnamefont {Liu}}, \bibinfo {author} {\bibfnamefont {J.}~\bibnamefont {Wurtz}}, \bibinfo {author} {\bibfnamefont {M.~D.}\ \bibnamefont {Lukin}}, \bibinfo {author} {\bibfnamefont {S.-T.}\ \bibnamefont {Wang}},\ and\ \bibinfo {author} {\bibfnamefont {H.}~\bibnamefont {Pichler}},\ }\href {https://doi.org/10.1103/PRXQuantum.4.010316} {\bibfield  {journal} {\bibinfo  {journal} {PRX Quantum}\ }\textbf {\bibinfo {volume} {4}},\ \bibinfo {pages} {010316} (\bibinfo {year} {2023})}\BibitemShut {NoStop}%
\bibitem [{\citenamefont {Farhi}\ \emph {et~al.}(2001)\citenamefont {Farhi}, \citenamefont {Goldstone}, \citenamefont {Gutmann}, \citenamefont {Lapan}, \citenamefont {Lundgren},\ and\ \citenamefont {Preda}}]{farhi-2001-quantum}%
  \BibitemOpen
  \bibfield  {author} {\bibinfo {author} {\bibfnamefont {E.}~\bibnamefont {Farhi}}, \bibinfo {author} {\bibfnamefont {J.}~\bibnamefont {Goldstone}}, \bibinfo {author} {\bibfnamefont {S.}~\bibnamefont {Gutmann}}, \bibinfo {author} {\bibfnamefont {J.}~\bibnamefont {Lapan}}, \bibinfo {author} {\bibfnamefont {A.}~\bibnamefont {Lundgren}},\ and\ \bibinfo {author} {\bibfnamefont {D.}~\bibnamefont {Preda}},\ }\href@noop {} {\bibfield  {journal} {\bibinfo  {journal} {Science}\ }\textbf {\bibinfo {volume} {292}},\ \bibinfo {pages} {472} (\bibinfo {year} {2001})}\BibitemShut {NoStop}%
\bibitem [{\citenamefont {Albash}\ and\ \citenamefont {Lidar}(2018)}]{albash-2018-adiabatic-computing}%
  \BibitemOpen
  \bibfield  {author} {\bibinfo {author} {\bibfnamefont {T.}~\bibnamefont {Albash}}\ and\ \bibinfo {author} {\bibfnamefont {D.~A.}\ \bibnamefont {Lidar}},\ }\href {https://doi.org/10.1103/RevModPhys.90.015002} {\bibfield  {journal} {\bibinfo  {journal} {Rev. Mod. Phys.}\ }\textbf {\bibinfo {volume} {90}},\ \bibinfo {pages} {015002} (\bibinfo {year} {2018})}\BibitemShut {NoStop}%
\bibitem [{\citenamefont {Schiffer}\ \emph {et~al.}(2024)\citenamefont {Schiffer}, \citenamefont {Wild}, \citenamefont {Maskara}, \citenamefont {Cain}, \citenamefont {Lukin},\ and\ \citenamefont {Samajdar}}]{Schiffer-2024-Circumventing-exp-runtimes}%
  \BibitemOpen
  \bibfield  {author} {\bibinfo {author} {\bibfnamefont {B.~F.}\ \bibnamefont {Schiffer}}, \bibinfo {author} {\bibfnamefont {D.~S.}\ \bibnamefont {Wild}}, \bibinfo {author} {\bibfnamefont {N.}~\bibnamefont {Maskara}}, \bibinfo {author} {\bibfnamefont {M.}~\bibnamefont {Cain}}, \bibinfo {author} {\bibfnamefont {M.~D.}\ \bibnamefont {Lukin}},\ and\ \bibinfo {author} {\bibfnamefont {R.}~\bibnamefont {Samajdar}},\ }\href {https://doi.org/10.1103/PhysRevResearch.6.013271} {\bibfield  {journal} {\bibinfo  {journal} {Phys. Rev. Res.}\ }\textbf {\bibinfo {volume} {6}},\ \bibinfo {pages} {013271} (\bibinfo {year} {2024})}\BibitemShut {NoStop}%
\bibitem [{\citenamefont {Miessen}\ \emph {et~al.}(2024)\citenamefont {Miessen}, \citenamefont {Egger}, \citenamefont {Tavernelli},\ and\ \citenamefont {Mazzola}}]{miessen-2024-defect-density-annealing}%
  \BibitemOpen
  \bibfield  {author} {\bibinfo {author} {\bibfnamefont {A.}~\bibnamefont {Miessen}}, \bibinfo {author} {\bibfnamefont {D.~J.}\ \bibnamefont {Egger}}, \bibinfo {author} {\bibfnamefont {I.}~\bibnamefont {Tavernelli}},\ and\ \bibinfo {author} {\bibfnamefont {G.}~\bibnamefont {Mazzola}},\ }\href@noop {} {\bibfield  {journal} {\bibinfo  {journal} {PRX Quantum}\ }\textbf {\bibinfo {volume} {5}},\ \bibinfo {pages} {040320} (\bibinfo {year} {2024})}\BibitemShut {NoStop}%
\bibitem [{\citenamefont {Bombieri}\ \emph {et~al.}(2024)\citenamefont {Bombieri}, \citenamefont {Zeng}, \citenamefont {Tricarico}, \citenamefont {Lin}, \citenamefont {Notarnicola}, \citenamefont {Cain}, \citenamefont {Lukin},\ and\ \citenamefont {Pichler}}]{bombieri-2024-CL-gap}%
  \BibitemOpen
  \bibfield  {author} {\bibinfo {author} {\bibfnamefont {L.}~\bibnamefont {Bombieri}}, \bibinfo {author} {\bibfnamefont {Z.}~\bibnamefont {Zeng}}, \bibinfo {author} {\bibfnamefont {R.}~\bibnamefont {Tricarico}}, \bibinfo {author} {\bibfnamefont {R.}~\bibnamefont {Lin}}, \bibinfo {author} {\bibfnamefont {S.}~\bibnamefont {Notarnicola}}, \bibinfo {author} {\bibfnamefont {M.}~\bibnamefont {Cain}}, \bibinfo {author} {\bibfnamefont {M.~D.}\ \bibnamefont {Lukin}},\ and\ \bibinfo {author} {\bibfnamefont {H.}~\bibnamefont {Pichler}},\ }\href@noop {} {\bibfield  {journal} {\bibinfo  {journal} {arXiv preprint arXiv:2411.04645}\ } (\bibinfo {year} {2024})}\BibitemShut {NoStop}%
\bibitem [{\citenamefont {Urban}\ \emph {et~al.}(2009)\citenamefont {Urban}, \citenamefont {Johnson}, \citenamefont {Henage}, \citenamefont {Isenhower}, \citenamefont {Yavuz}, \citenamefont {Walker},\ and\ \citenamefont {Saffman}}]{urban-2009-rydberg-blockade}%
  \BibitemOpen
  \bibfield  {author} {\bibinfo {author} {\bibfnamefont {E.}~\bibnamefont {Urban}}, \bibinfo {author} {\bibfnamefont {T.~A.}\ \bibnamefont {Johnson}}, \bibinfo {author} {\bibfnamefont {T.}~\bibnamefont {Henage}}, \bibinfo {author} {\bibfnamefont {L.}~\bibnamefont {Isenhower}}, \bibinfo {author} {\bibfnamefont {D.}~\bibnamefont {Yavuz}}, \bibinfo {author} {\bibfnamefont {T.}~\bibnamefont {Walker}},\ and\ \bibinfo {author} {\bibfnamefont {M.}~\bibnamefont {Saffman}},\ }\href@noop {} {\bibfield  {journal} {\bibinfo  {journal} {Nature Physics}\ }\textbf {\bibinfo {volume} {5}},\ \bibinfo {pages} {110} (\bibinfo {year} {2009})}\BibitemShut {NoStop}%
\bibitem [{sup()}]{supp}%
  \BibitemOpen
  \href@noop {} {}\bibinfo {note} {See Supplemental Material at URL-will-be-inserted-by-publisher for the data of the experiments.}\BibitemShut {Stop}%
\bibitem [{\citenamefont {Liu}\ \emph {et~al.}(2023)\citenamefont {Liu}, \citenamefont {Gao}, \citenamefont {Cain}, \citenamefont {Lukin},\ and\ \citenamefont {Wang}}]{Liu-2023-GNT}%
  \BibitemOpen
  \bibfield  {author} {\bibinfo {author} {\bibfnamefont {J.-G.}\ \bibnamefont {Liu}}, \bibinfo {author} {\bibfnamefont {X.}~\bibnamefont {Gao}}, \bibinfo {author} {\bibfnamefont {M.}~\bibnamefont {Cain}}, \bibinfo {author} {\bibfnamefont {M.~D.}\ \bibnamefont {Lukin}},\ and\ \bibinfo {author} {\bibfnamefont {S.-T.}\ \bibnamefont {Wang}},\ }\href {https://doi.org/10.1137/22m1501787} {\bibfield  {journal} {\bibinfo  {journal} {SIAM Journal on Scientific Computing}\ }\textbf {\bibinfo {volume} {45}},\ \bibinfo {pages} {A1239–A1270} (\bibinfo {year} {2023})}\BibitemShut {NoStop}%
\bibitem [{\citenamefont {Levit}\ and\ \citenamefont {Mandrescu}(2003)}]{Levit-Mandrescu-2003}%
  \BibitemOpen
  \bibfield  {author} {\bibinfo {author} {\bibfnamefont {V.~E.}\ \bibnamefont {Levit}}\ and\ \bibinfo {author} {\bibfnamefont {E.}~\bibnamefont {Mandrescu}},\ }in\ \href@noop {} {\emph {\bibinfo {booktitle} {Discrete Mathematics and Theoretical Computer Science}}},\ \bibinfo {editor} {edited by\ \bibinfo {editor} {\bibfnamefont {C.~S.}\ \bibnamefont {Calude}}, \bibinfo {editor} {\bibfnamefont {M.~J.}\ \bibnamefont {Dinneen}},\ and\ \bibinfo {editor} {\bibfnamefont {V.}~\bibnamefont {Vajnovszki}}}\ (\bibinfo  {publisher} {Springer Berlin Heidelberg},\ \bibinfo {address} {Berlin, Heidelberg},\ \bibinfo {year} {2003})\ pp.\ \bibinfo {pages} {237--256}\BibitemShut {NoStop}%
\bibitem [{\citenamefont {Arocha}(1984)}]{arocha-1984-propriedades}%
  \BibitemOpen
  \bibfield  {author} {\bibinfo {author} {\bibfnamefont {J.~L.}\ \bibnamefont {Arocha}},\ }\href@noop {} {\bibfield  {journal} {\bibinfo  {journal} {Revista Ciencias Matematicas}\ }\textbf {\bibinfo {volume} {5}},\ \bibinfo {pages} {103} (\bibinfo {year} {1984})}\BibitemShut {NoStop}%
\bibitem [{\citenamefont {Gilbert}(1959)}]{gilbert-1959-random-graphs}%
  \BibitemOpen
  \bibfield  {author} {\bibinfo {author} {\bibfnamefont {E.~N.}\ \bibnamefont {Gilbert}},\ }\href {https://doi.org/10.1214/aoms/1177706098} {\bibfield  {journal} {\bibinfo  {journal} {The Annals of Mathematical Statistics}\ }\textbf {\bibinfo {volume} {30}},\ \bibinfo {pages} {1141 } (\bibinfo {year} {1959})}\BibitemShut {NoStop}%
\bibitem [{\citenamefont {de~Correc}(2025)}]{correc-2025}%
  \BibitemOpen
  \bibfield  {author} {\bibinfo {author} {\bibfnamefont {C.}~\bibnamefont {de~Correc}},\ }\href {https://doi.org/10.5281/zenodo.1234567} {10.5281/zenodo.1234567} (\bibinfo {year} {2025})\BibitemShut {NoStop}%
\bibitem [{\citenamefont {Dalyac}\ \emph {et~al.}(2023)\citenamefont {Dalyac}, \citenamefont {Henry}, \citenamefont {Kim}, \citenamefont {Ahn},\ and\ \citenamefont {Henriet}}]{dalyac-2023-exploring-graph-locality}%
  \BibitemOpen
  \bibfield  {author} {\bibinfo {author} {\bibfnamefont {C.}~\bibnamefont {Dalyac}}, \bibinfo {author} {\bibfnamefont {L.-P.}\ \bibnamefont {Henry}}, \bibinfo {author} {\bibfnamefont {M.}~\bibnamefont {Kim}}, \bibinfo {author} {\bibfnamefont {J.}~\bibnamefont {Ahn}},\ and\ \bibinfo {author} {\bibfnamefont {L.}~\bibnamefont {Henriet}},\ }\href {https://doi.org/10.1103/PhysRevA.108.052423} {\bibfield  {journal} {\bibinfo  {journal} {Phys. Rev. A}\ }\textbf {\bibinfo {volume} {108}},\ \bibinfo {pages} {052423} (\bibinfo {year} {2023})}\BibitemShut {NoStop}%
\bibitem [{\citenamefont {Lechner}\ \emph {et~al.}(2015)\citenamefont {Lechner}, \citenamefont {Hauke},\ and\ \citenamefont {Zoller}}]{lechner-2015-LHZ}%
  \BibitemOpen
  \bibfield  {author} {\bibinfo {author} {\bibfnamefont {W.}~\bibnamefont {Lechner}}, \bibinfo {author} {\bibfnamefont {P.}~\bibnamefont {Hauke}},\ and\ \bibinfo {author} {\bibfnamefont {P.}~\bibnamefont {Zoller}},\ }\href@noop {} {\bibfield  {journal} {\bibinfo  {journal} {Science advances}\ }\textbf {\bibinfo {volume} {1}},\ \bibinfo {pages} {e1500838} (\bibinfo {year} {2015})}\BibitemShut {NoStop}%
\bibitem [{\citenamefont {Lanthaler}\ \emph {et~al.}(2023)\citenamefont {Lanthaler}, \citenamefont {Dlaska}, \citenamefont {Ender},\ and\ \citenamefont {Lechner}}]{lanthaler-2023-LHZ}%
  \BibitemOpen
  \bibfield  {author} {\bibinfo {author} {\bibfnamefont {M.}~\bibnamefont {Lanthaler}}, \bibinfo {author} {\bibfnamefont {C.}~\bibnamefont {Dlaska}}, \bibinfo {author} {\bibfnamefont {K.}~\bibnamefont {Ender}},\ and\ \bibinfo {author} {\bibfnamefont {W.}~\bibnamefont {Lechner}},\ }\href@noop {} {\bibfield  {journal} {\bibinfo  {journal} {Physical Review Letters}\ }\textbf {\bibinfo {volume} {130}},\ \bibinfo {pages} {220601} (\bibinfo {year} {2023})}\BibitemShut {NoStop}%
\bibitem [{\citenamefont {Kibble}(1976)}]{kibble-1976-cosmo}%
  \BibitemOpen
  \bibfield  {author} {\bibinfo {author} {\bibfnamefont {T.~W.}\ \bibnamefont {Kibble}},\ }\href@noop {} {\bibfield  {journal} {\bibinfo  {journal} {Journal of Physics A: Mathematical and General}\ }\textbf {\bibinfo {volume} {9}},\ \bibinfo {pages} {1387} (\bibinfo {year} {1976})}\BibitemShut {NoStop}%
\bibitem [{\citenamefont {Zurek}(1985)}]{zurek-1985-cosmological}%
  \BibitemOpen
  \bibfield  {author} {\bibinfo {author} {\bibfnamefont {W.~H.}\ \bibnamefont {Zurek}},\ }\href@noop {} {\bibfield  {journal} {\bibinfo  {journal} {Nature}\ }\textbf {\bibinfo {volume} {317}},\ \bibinfo {pages} {505} (\bibinfo {year} {1985})}\BibitemShut {NoStop}%
\bibitem [{\citenamefont {Zurek}(1996)}]{zurek-1996-cosmological}%
  \BibitemOpen
  \bibfield  {author} {\bibinfo {author} {\bibfnamefont {W.~H.}\ \bibnamefont {Zurek}},\ }\href@noop {} {\bibfield  {journal} {\bibinfo  {journal} {Physics Reports}\ }\textbf {\bibinfo {volume} {276}},\ \bibinfo {pages} {177} (\bibinfo {year} {1996})}\BibitemShut {NoStop}%
\bibitem [{\citenamefont {Bunkov}\ and\ \citenamefont {Godfrin}(2000)}]{bunkov-2000-topological-defects}%
  \BibitemOpen
  \bibfield  {author} {\bibinfo {author} {\bibfnamefont {Y.~M.}\ \bibnamefont {Bunkov}}\ and\ \bibinfo {author} {\bibfnamefont {H.}~\bibnamefont {Godfrin}},\ }\href@noop {} {\emph {\bibinfo {title} {Topological defects and the non-equilibrium dynamics of symmetry breaking phase transitions}}},\ Vol.\ \bibinfo {volume} {549}\ (\bibinfo  {publisher} {Springer Science \& Business Media},\ \bibinfo {year} {2000})\BibitemShut {NoStop}%
\bibitem [{\citenamefont {Caro}(1979)}]{caro-1979-wei-theorem}%
  \BibitemOpen
  \bibfield  {author} {\bibinfo {author} {\bibfnamefont {Y.}~\bibnamefont {Caro}},\ }\href@noop {} {\emph {\bibinfo {title} {New results on the independence number}}},\ \bibinfo {type} {Tech. Rep.}\ (\bibinfo  {institution} {Technical Report, Tel-Aviv University},\ \bibinfo {year} {1979})\BibitemShut {NoStop}%
\bibitem [{\citenamefont {Wei}(1981)}]{wei-1981-caro-theorem}%
  \BibitemOpen
  \bibfield  {author} {\bibinfo {author} {\bibfnamefont {V.~K.}\ \bibnamefont {Wei}},\ }\href@noop {} {\bibinfo {title} {A lower bound on the stability number of a simple graph}} (\bibinfo {year} {1981})\BibitemShut {NoStop}%
\bibitem [{\citenamefont {Coja-Oghlan}\ and\ \citenamefont {Efthymiou}(2011)}]{coja-2011}%
  \BibitemOpen
  \bibfield  {author} {\bibinfo {author} {\bibfnamefont {A.}~\bibnamefont {Coja-Oghlan}}\ and\ \bibinfo {author} {\bibfnamefont {C.}~\bibnamefont {Efthymiou}},\ }in\ \href@noop {} {\emph {\bibinfo {booktitle} {Society for Industrial and Applied Mathematics and Association for Computing Machinery. Proceeding of the ACM-SIAM Symposium on Discrete Algorithms}}}\ (\bibinfo {organization} {Society for Industrial and Applied Mathematics},\ \bibinfo {year} {2011})\ p.\ \bibinfo {pages} {136}\BibitemShut {NoStop}%
\bibitem [{\citenamefont {Bazgan}\ \emph {et~al.}(2005)\citenamefont {Bazgan}, \citenamefont {Escoffier},\ and\ \citenamefont {Paschos}}]{bazgan-2005-approximating-MIS-is-hard}%
  \BibitemOpen
  \bibfield  {author} {\bibinfo {author} {\bibfnamefont {C.}~\bibnamefont {Bazgan}}, \bibinfo {author} {\bibfnamefont {B.}~\bibnamefont {Escoffier}},\ and\ \bibinfo {author} {\bibfnamefont {V.~T.}\ \bibnamefont {Paschos}},\ }\href@noop {} {\bibfield  {journal} {\bibinfo  {journal} {Theoretical Computer Science}\ }\textbf {\bibinfo {volume} {339}},\ \bibinfo {pages} {272} (\bibinfo {year} {2005})}\BibitemShut {NoStop}%
\bibitem [{\citenamefont {Aharonov}\ \emph {et~al.}(2013)\citenamefont {Aharonov}, \citenamefont {Arad},\ and\ \citenamefont {Vidick}}]{aharonov-2013-quantum-pcp}%
  \BibitemOpen
  \bibfield  {author} {\bibinfo {author} {\bibfnamefont {D.}~\bibnamefont {Aharonov}}, \bibinfo {author} {\bibfnamefont {I.}~\bibnamefont {Arad}},\ and\ \bibinfo {author} {\bibfnamefont {T.}~\bibnamefont {Vidick}},\ }\href {https://arxiv.org/abs/1309.7495} {\bibinfo {title} {The quantum pcp conjecture}} (\bibinfo {year} {2013}),\ \Eprint {https://arxiv.org/abs/1309.7495} {arXiv:1309.7495 [quant-ph]} \BibitemShut {NoStop}%
\bibitem [{\citenamefont {Park}\ \emph {et~al.}(2024)\citenamefont {Park}, \citenamefont {Jeong}, \citenamefont {Kim}, \citenamefont {Kim}, \citenamefont {Byun}, \citenamefont {Vignoli}, \citenamefont {Henry}, \citenamefont {Henriet},\ and\ \citenamefont {Ahn}}]{park-2024-factorization-with-rydberg}%
  \BibitemOpen
  \bibfield  {author} {\bibinfo {author} {\bibfnamefont {J.}~\bibnamefont {Park}}, \bibinfo {author} {\bibfnamefont {S.}~\bibnamefont {Jeong}}, \bibinfo {author} {\bibfnamefont {M.}~\bibnamefont {Kim}}, \bibinfo {author} {\bibfnamefont {K.}~\bibnamefont {Kim}}, \bibinfo {author} {\bibfnamefont {A.}~\bibnamefont {Byun}}, \bibinfo {author} {\bibfnamefont {L.}~\bibnamefont {Vignoli}}, \bibinfo {author} {\bibfnamefont {L.-P.}\ \bibnamefont {Henry}}, \bibinfo {author} {\bibfnamefont {L.}~\bibnamefont {Henriet}},\ and\ \bibinfo {author} {\bibfnamefont {J.}~\bibnamefont {Ahn}},\ }\href@noop {} {\bibfield  {journal} {\bibinfo  {journal} {Physical Review Research}\ }\textbf {\bibinfo {volume} {6}},\ \bibinfo {pages} {023241} (\bibinfo {year} {2024})}\BibitemShut {NoStop}%
\bibitem [{\citenamefont {Byun}\ \emph {et~al.}(2024)\citenamefont {Byun}, \citenamefont {Jung}, \citenamefont {Kim}, \citenamefont {Kim}, \citenamefont {Jeong}, \citenamefont {Jeong},\ and\ \citenamefont {Ahn}}]{byun-2024-QUBO-with-Rydberg}%
  \BibitemOpen
  \bibfield  {author} {\bibinfo {author} {\bibfnamefont {A.}~\bibnamefont {Byun}}, \bibinfo {author} {\bibfnamefont {J.}~\bibnamefont {Jung}}, \bibinfo {author} {\bibfnamefont {K.}~\bibnamefont {Kim}}, \bibinfo {author} {\bibfnamefont {M.}~\bibnamefont {Kim}}, \bibinfo {author} {\bibfnamefont {S.}~\bibnamefont {Jeong}}, \bibinfo {author} {\bibfnamefont {H.}~\bibnamefont {Jeong}},\ and\ \bibinfo {author} {\bibfnamefont {J.}~\bibnamefont {Ahn}},\ }\href@noop {} {\bibfield  {journal} {\bibinfo  {journal} {Advanced Quantum Technologies}\ }\textbf {\bibinfo {volume} {7}},\ \bibinfo {pages} {2300398} (\bibinfo {year} {2024})}\BibitemShut {NoStop}%
\end{thebibliography}%
\end{document}